\documentclass[acmsmall]{acmart}

\setcopyright{acmlicensed}
\acmJournal{FACMP}  
\acmYear{2023}
\acmVolume{}
\acmNumber{}
\acmArticle{}
\acmMonth{4}
\acmPrice{15.00}
\acmDOI{10.1145/3592624}

\title{Decentralized Inverse Transparency With Blockchain}

\newcommand{\vzaffil}{%
	\institution{Technical University of Munich}
	\city{Munich}
	\country{Germany}%
}

\author{Valentin Zieglmeier}
\orcid{0000-0002-3770-0321}
\affiliation{\vzaffil{}}
\email{valentin.zieglmeier@tum.de}

\author{Gabriel Loyola Daiqui}
\orcid{0000-0002-9961-0866}
\affiliation{\vzaffil{}}
\email{gabriel.loyola-daiqui@tum.de}

\author{Alexander Pretschner}
\orcid{0000-0002-5573-1201}
\affiliation{\vzaffil{}}
\email{alexander.pretschner@tum.de}

\authorsaddresses{%
	Authors' addresses: Valentin Zieglmeier, valentin.zieglmeier@tum.de; Gabriel Loyola Daiqui, gabriel.loyola-daiqui@tum.de; Alexander Pretschner, alexander.pretschner@tum.de, Technical University of Munich, TUM School of Computation, Information and Technology, Chair of Software and Systems Engineering, Boltzmannstr. 3, 85748 Garching, Germany%
}

\keywords{Decentralization, Non-repudiation, Accountability, Privacy, Anonymity}

\usepackage{subcaption}

\newcommand{\reqref}[1]{(\ref{#1})}
\newcommand{\stepreq}[1]{%
	\refstepcounter{Requirement}%
	\label{#1}%
}
\newcommand{\newreq}[1]{%
	\stepreq{#1}%
	\reqref{#1}%
}
\newcommand{\PPP}{\textsc{P\textsuperscript{3}}}
\newcommand{\Kovacs}{\textsc{Kovacs}}

\begin{document}

	\begin{abstract}

		Employee data can be used to facilitate work, but their misusage may pose risks for individuals.
		\emph{Inverse transparency} therefore aims to track all usages of personal data, allowing individuals to monitor them to ensure accountability for potential misusage.
		This necessitates a trusted log to establish an agreed-upon and non-repudiable timeline of events.
		The unique properties of blockchain facilitate this by providing immutability and availability.
		For power asymmetric environments such as the workplace, permissionless blockchain is especially beneficial as no trusted third party is required.
		Yet, two issues remain:
		(1) In a decentralized environment, no arbiter can facilitate and attest to data exchanges.
		Simple peer-to-peer sharing of data, conversely, lacks the required non-repudiation.
		(2) With data governed by privacy legislation such as the GDPR, the core advantage of immutability becomes a liability.
		After a rightful request, an individual's personal data need to be rectified or deleted, which is impossible in an immutable blockchain.

		To solve these issues, we present \Kovacs{}, a decentralized data exchange and usage logging system for inverse transparency built on blockchain.
		Its new-usage protocol ensures non-repudiation, and therefore accountability, for inverse transparency.
		Its one-time pseudonym generation algorithm guarantees unlinkability and enables proof of ownership, which allows data subjects to exercise their legal rights regarding their personal data.
		With our implementation, we show the viability of our solution.
		The decentralized communication impacts performance and scalability, but exchange duration and storage size are still reasonable.
		More importantly, the provided information security meets high requirements.
		We conclude that \Kovacs{} realizes decentralized inverse transparency through secure and GDPR-compliant use of permissionless blockchain.

	\end{abstract}

	\ccsdesc[500]{Computer systems organization~Peer-to-peer architectures}
	\ccsdesc[500]{Security and privacy~Distributed systems security}
	\ccsdesc[300]{Security and privacy~Privacy-preserving protocols}
	\ccsdesc[300]{Security and privacy~Cryptography}

	\maketitle

	\section{Introduction}

Employee data collected in the workplace can be a valuable source for analyses and predictions.
So-called \emph{people analytics} tools utilize these data to help improve collaboration and facilitate work~\cite{tursunbayeva2018people}.
Yet, advanced analytics also increase the risk of misinterpretations or data misusage~\cite{tursunbayeva2021ethics}.
To protect employees from malicious usage of their data, the concept of \emph{inverse transparency}~\cite{brin1998transparent} has been introduced to the workplace.
That entails that all usages of personal data in people analytics are tracked, stored in a tamper-proof log, and made available to the data owners~\cite{zieglmeier2023itbd}.
This allows individuals more oversight and control in situations of asymmetric power, such as the workplace.
For such a usage log to enable accountability, one needs to establish an agreed-upon and non-repudiable timeline of events~\cite{zhou1996observations} and guarantee its integrity~\cite{schaefer2019transparent, kelbert2018data}.
More importantly, no single party can be trusted with managing this log due to the inherent power asymmetry in the workplace.
Otherwise, manipulation of the logs by, e.g., removing incriminating evidence would be possible, preventing accountability.
To achieve this, \citeauthor{schaefer2019transparent} recently proposed utilizing blockchain for a secure usage log~\cite{schaefer2019transparent}.
Blockchain can offer advantages in contexts with untrusted participants, especially if immutability of data is required~\cite{zheng2017overview, taylor2020systematic}.
Consequently, multiple other secure logs based on blockchain were developed in recent years~\cite[e.g.,][]{ge2019permission, shekhtman2021engravechain}.
The technology has many advantages for these applications, as it is an effective way to guarantee immutability of stored entries (integrity) and functions even in unreliable distributed networks (availability).

Yet, in the context of data sharing, permissionless blockchain has two core limitations.
First, with no trusted third party, no single arbiter can attest to the successful completion of data exchanges.
Both sides of an exchange have an incentive to lie; the recipient of data may claim that they never received the data, while the sender may not send it but claim that they did.
To guarantee integrity of the usage log, we therefore require non-repudiable data exchanges.
Second, the unique properties of blockchain mean that the confidentiality of stored data cannot be guaranteed by default.
Even when storing minimal information, some form of identifier is required to denote ownership or association to entries. Without necessitating any information leak, the blockchain then allows any network participant to trace entries based on their identifier~\cite{reid2013analysis}.
At best, users can try to hide their association to blockchain entries by keeping their identifier secret and creating new addresses.
Even then, network participants can deduce information about users simply by analyzing publicly available data~\cite{androulaki2013evaluating}.
If the identifier is leaked or known to a third party, though, all respective entries can be retroactively associated with them.
This has been identified as a problem of blockchain-based secure logs~\cite{ge2019permission, schaefer2019transparent}, where confidentiality can be an important property~\cite{accorsi2010bbox}.
More critically, recent privacy legislation such as the General Data Protection Regulation (GDPR)~\cite{eu2016gdpr} of the European Union and the California Consumer Privacy Act (CCPA)~\cite{cali2018privacy} requires those who store personal data to protect and, on request, even delete it.
As data stored in blockchain can be identifiable, even with typical protection measures, they fall under the provisions of privacy legislation.
Especially the right to erasure has been identified as a core issue of blockchain in this context~\cite{pagallo2018chronicle}.

Intuitively, it seems as if this means a fundamental conflict between the requirements of inverse transparency and privacy legislation on the one hand, and permissionless blockchain technology on the other hand.
Therefore, the only solution would be not to utilize blockchain when providing inverse transparency.
We argue that this conflict can be solved differently, though.
We aim to combine the strengths of blockchain (such as decentralization and immutability) with the requirements of inverse transparency (accountability) and privacy legislation (confidentiality, deletability).

\paragraph{Contribution:} Blockchain is uniquely positioned to solve many issues of inverse transparency in a decentralized environment.
Yet, the goal of accountability for data usages requires a solution that guarantees non-repudiable data exchanges.
Furthermore, secure usage logs based on blockchain are, by default, fundamentally at odds with privacy legislation such as the GDPR.
The metadata recorded in blockchain are themselves personal data that need to be protected and, on request, rectified or deleted.
To tackle these challenges and enable decentralized inverse transparency, we therefore contribute the data exchange and blockchain logging system \Kovacs{} with its core components, the new-usage protocol and private pseudonym provisioning.
Our contribution encompasses its concept and algorithms as well as a complete open-source implementation.
The new-usage protocol enables secure and decentralized data exchange while ensuring non-repudiation, as required for accountability.
The one-time pseudonym generation algorithm, meanwhile, guarantees two properties:
Proof of ownership, as required for deletion requests, and unlinkability, to provide users anonymity against adversaries.
Notably, \Kovacs{} does not require any changes in the utilized blockchain software and can therefore even run on arbitrary public blockchains.

\paragraph{Extension:} This paper is an extended version of our previously published short paper~\cite{zieglmeier2021gdpr}.
In it, we introduced the \PPP{} concept, encompassing a new-usage protocol and private pseudonym provisioning, and analyzed its security properties theoretically.
We extend upon that in multiple, significant ways:
(1)~We present and implement \Kovacs{}, a complete data exchange and usage log blockchain system that integrates \PPP{} to increase information security and GDPR compatibility.
To that end, we add a refined use case (Sec.~\ref{sec:background}) motivating our work, expand on the requirements, adversarial model, and system concept (Sec.~\ref{sec:concept}), add an implementation (Sec.~\ref{sec:implementation}), and expand our discussion of related works (Sec.~\ref{sec:related-work}) to cover non-repudiable data exchange.
(2)~We improve the new-usage protocol (Sec.~\ref{sec:generation-protocol}) to simplify its implementation without compromising security.
(3)~In addition to our theoretical analyses (Secs.~\ref{sec:evaluation-security}--\ref{sec:evaluation-gdpr}), we add an evaluation of the performance and scalability of the implemented \Kovacs{} instance (Secs.~\ref{sec:evaluation:performance}--\ref{sec:evaluation:scalability}).

	\section{Background} \label{sec:background}

We first go into more detail regarding the concept of \emph{inverse transparency} and why we think its current realization is in need of decentralization.
Then, we outline the legal difference between pseudonymity and anonymity, an important detail that we make use of later.

\subsection{Inverse Transparency} \label{sec:theory:inverse-transparency}

Typically, when personal data are handled, their usage is covered by privacy policies or company agreements. These policies are hard to read and understand~\cite{mcdonald2008cost}, calling into question whether individuals subjected to them truly understand their impact.
Especially in the workplace, this can become problematic.
While some usages of their data might be beneficial for employees, giving access to personal data poses the risk of profiling and misusage.
The inherent power asymmetry and forced technology adoption exacerbate these risks~\cite{tursunbayeva2021ethics, zieglmeier2022increasing}.
To give employees more oversight and control in this situation of asymmetric knowledge and power, the concept of inverse transparency~\cite{brin1998transparent} was introduced to the workplace.
At its core, it is based on the principle that access to personal data should be visible (transparent) to data owners~\cite{brin1998transparent}.
\citeauthor{gierlich2020more} initially described how this idea can be applied abstractly as a digital leadership concept~\cite{gierlich2020more}.
To realize inverse transparency technically, \citeauthor{zieglmeier2023itbd} propose to design people analytics software from the ground up to track the flow of data and create a data usage log~\cite{zieglmeier2023itbd}.
Their \emph{inverse transparency toolchain} is inherently centralized, though.
It requires trust in multiple parties, such as the employer and the system administrators~\cite[see][]{zieglmeier2021trustworthy, zieglmeier2023toolchain}.

Tracking all data usages is an important prerequisite for inverse transparency, but for true accountability we need to be able to guarantee completeness and correctness of the created usage log.
Due to the inherent power asymmetry in the workplace, it is in our view not sufficient to consider the employer a trusted party that can safeguard the logs.
Ideally, no trusted third party is required at all, as they might be interested to modify the log and, e.g., remove potentially incriminating evidence.
Therefore, we consider it necessary to use a distributed, tamper-proof logging mechanism to guarantee accountability~\cite[see, e.g.,][]{accorsi2010bbox, zyskind2015decentralizing, schaefer2019transparent}.

In the following, we refer to the participants in a data sharing transaction as the \emph{data owner} and the \emph{data consumer}.
The \emph{data owner} ``possesses the rights to the data''~\cite[p.~40]{pretschner2006distributed}. The GDPR refers to them as the ``data subject''.
The \emph{data consumer}, meanwhile, is the person or program that processes, and thereby ``consumes'', personal data that identify one or more \emph{data owners}.~\cite[p.~40]{pretschner2006distributed};~\cite{zieglmeier2023itbd}

\subsection{Pseudonymity and Anonymity}\label{sec:pseudonymity-anonymity}

Two concepts are important to understand when discussing the applicability and implications of the GDPR: pseudonymity and anonymity.

Pseudonymized data are personal data where identifiers (such as names) have been replaced by pseudonyms, and the association between pseudonyms and identifiers is stored separately from the data themselves.
As the availability of this link allows re-identification, these data are not anonymous and fall under the provisions of privacy legislation~\cite{enisa2019recommendations}.
Anonymized data on the other hand are data that have been modified as such to make it impossible to re-identify an individual from them~\cite{iso25237pseudonymization, enisa2019recommendations}.
Importantly: While pseudonymous data are regarded personal data, anonymous data are not~\cite{enisa2019recommendations}. This means that to fulfill a user's legal right to erasure, we do not actually need to delete their personal data, as long as we can anonymize it by irreversibly deleting all existing links to the pseudonym~\cite[p.~153]{hintze2018comparing}.
	\section{Concept: Decentralized Inverse Transparency With Blockchain} \label{sec:concept}

We present \Kovacs{}, an inverse transparency log system that encompasses two core parts: non-repudiable data exchange and private pseudonym provisioning.
Inverse transparency requires accountability for occurred data usages, which is why we develop a non-repudiable data exchange protocol.
Evidences for the exchange are stored in a blockchain, ensuring log integrity.
To protect confidentiality of the stored data while preserving proof of ownership, we present a pseudonym provisioning algorithm as the foundational differentiator of our concept.

First, we define our adversarial model and attacks that we consider. From our use case and the adversarial model, we derive requirements.
Finally, we detail the \Kovacs{} system, a complete solution for decentralized inverse transparency with blockchain.

\subsection{Adversarial Model}\label{sec:adversarial-model}

A user $u$ may be either in the set of \emph{data owners} $O = \{o_1,o_2,...,o_n\}$, in the set of \emph{data consumers} $C = \{c_1,c_2,...,c_m\}$, or both.
The adversary $\alpha$ can be any user and assume any role.
Whenever a consumer $c_i \in C$ accesses data of an owner $o_j \in O$, a usage $u_{ij}(c_i \to o_j)$ is appended to the usage log $U$ stored in the blockchain.

We assume that $\alpha$ has limited computational capacity, and can therefore never assume control over the blockchain network. Yet, within their means, they aim for maximum damage and therefore do not ``play fair''.
To start with, $\alpha$ aims to attack the integrity of the log.
As we use blockchain, preventing usages from being appended or retroactively modifying them is infeasible for $\alpha$~\cite{zheng2017overview}.
Therefore, they instead try to repudiate occurred data usages or claim fake ones.
Furthermore and more critically for blockchain, though, $\alpha$ is motivated to attack the confidentiality of the stored data by gaining access to information that is not meant to be accessible by them.

Specifically, $\alpha$ tries to conduct the following attacks:
\begin{enumerate}
	\renewcommand{\theenumi}{\alph{enumi}}
	\item \emph{Repudiate} a data usage $u_{ij}(c_i \to o_j)$.
	\item \emph{Fabricate} an entry $u_{ij}(c_i \to o_j)$ for a usage that did not occur.
	\item Derive from any entry $u_{ij}(c_i \to o_j)$ with $\alpha \notin \{c_i, o_j\}$ the identity of $c_i$ or $o_j$.
	\item Associate any two entries $u_{ij}(c_i \to o_j), u_{ik}(c_i \to o_k)$ with each other, thereby leaking their association with a single \emph{data consumer} $c_i$.
	\item Associate any two entries $u_{ji}(c_j \to o_i), u_{ki}(c_k \to o_i)$ with each other, thereby leaking their association with a single \emph{data owner} $o_i$.
	\item Leak the identity of $c_i$ for a stored usage $u_{ij}(c_i \to o_j)$ with $\alpha = o_j$, \emph{after} $c_i$ has legitimately exercised their right to erasure under the GDPR regarding $u_{ij}$.
\end{enumerate}

\subsection{Requirements} \label{sec:concept-requirements}
\newcounter{Requirement}

From our use case and adversarial model arise six main requirements:
\newreq{req:no-trust}~No trusted third party is necessary.
\newreq{req:non-repudiable}~The usage log needs to be non-repudiable and tamper-proof, preventing, e.g., data consumers from removing incriminating entries.
\newreq{req:non-forgeable}~The usage log needs to be non-forgeable, preventing, e.g., data owners from creating valid, but fabricated entries.
\newreq{req:owners-verify}~\emph{Data owners} can efficiently query for arbitrary log entries concerning their data and view their content. Importantly, they can prove the association of the data consumer to the logged usage (non-repudiation).
\newreq{req:consumers-verify}~\emph{Data consumers} can efficiently query for arbitrary log entries concerning their usages and verify their content.
\newreq{req:confidentiality}~No third party can derive identities or usage information from data in the blockchain.

\stepreq{req:gdpr}
In addition, the data stored in the usage log are governed by the GDPR for as long as they can be associated with the identities of users.
From that follows an additional requirement, namely compliance with the GDPR rights~\cite[Ch.~3]{eu2016gdpr}.
According to \citeauthor{pagallo2018chronicle} and \citeauthor{godyn2022analysis}, the main issue to solve for GDPR-compliant blockchain is the \emph{right to erasure}~\cite[Art.~17]{eu2016gdpr};~\cite{pagallo2018chronicle, godyn2022analysis}.
This is confirmed in the legal analysis of \citeauthor{tatar2020law}, who additionally note that a responsible controller may need to be identifiable to exercise that right~\cite{tatar2020law}.
We agree with their view, as we elaborate in the following.
Most GDPR rights, such as the \emph{right of access}~\cite[Art.~15]{eu2016gdpr}, are not negatively affected when storing personal data in a blockchain.
In fact, some may even be strengthened, as portability of and access to the data is enabled by design~\cite{jambert2019blockchain}.
Technical challenges only arise from the inherent immutability of blockchain.
This property implies that stored data cannot be deleted or altered, which affects the \emph{right to erasure} and the \emph{right to rectification}~\cite[Art.~16]{eu2016gdpr}.
We can generalize both issues to a single technical requirement, as enabling the deletion of personal data indirectly enables rectification: by removing the incorrect entry and adding the rectified version.
Furthermore, as we have discussed in Sec.~\ref{sec:pseudonymity-anonymity}, we can fulfill a user's legal right to erasure by anonymizing their personal data~\cite[p.~153]{hintze2018comparing}. Therefore, we arrive at requirement \reqref{req:gdpr}: The usage logs stored in the blockchain can be anonymized retroactively, making re-identification technically impossible.

\subsection{The \Kovacs{} System}

Our concept for the \Kovacs{} system consists of four parts: The new-usage protocol, the pseudonym generation algorithm, the block structure, and the deployment model.
First, the new-usage protocol guides the decentralized, non-repudiable communication between data consumer and data owner when data are shared and the usage is logged.
Second, the pseudonym generation algorithm enables the provisioning of private pseudonyms that guarantee unlinkability and proof of ownership.
Third, the block structure describes how usage data are stored in the blockchain and how they are protected.
Finally, the deployment model determines the required trust and computational resources.

\subsubsection{New-Usage Protocol}\label{sec:generation-protocol}

Many properties we aim for hinge on the specific protocol that is followed when a data usage occurs.
Concretely, that means a data consumer $c_i$ is accessing a datum $d$ of a data owner $o_j$ and this usage being logged in the blockchain.
This protocol is the first core step towards our goal.
The most important challenge hereby is to guarantee non-repudiation of the occurred usage without a trusted third party. Therefore, we adapt the non-repudiation protocol designed by \citeauthor{markowitch1999probabilistic}~\cite{markowitch1999probabilistic} for our use case.
In short: After the protocol is successfully completed, each party will possess proof of their interaction with the other party. Importantly, $o_j$ can prove that $c_i$ has received the datum $d$.

\begin{figure}[htbp]
	\centering
	\includegraphics[width=0.85\linewidth]{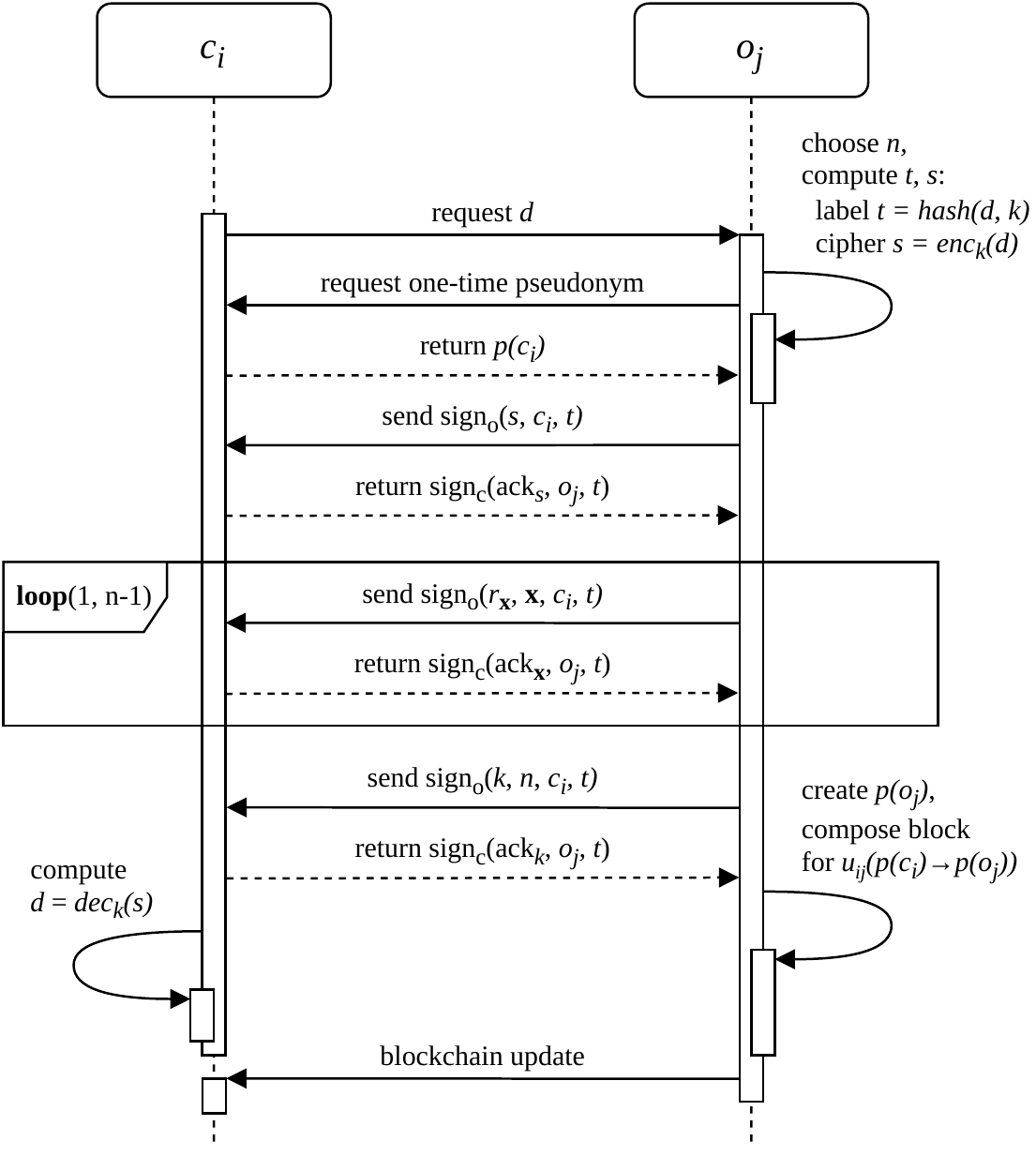}
	\caption{
		The new-usage protocol, adapted from~\cite{shamir1979share, markowitch1999probabilistic, kremer2002intensive}. $c_i$ requests access to the datum $d$ from $u_j$.
		After receiving pseudonym $p(c_i)$, $o_j$ sends the cipher $s$ and the label $t$, which is a hash of datum and key serving as an identifier for the transaction.
		The receipt of $s$ is acknowledged by $c_i$.
		Then, the iterative non-repudiation protocol is performed, with a random number of steps $n$ unknown to $c_i$~\cite[see][pp.~5--6]{markowitch1999probabilistic}.
		In each step $x$, $o_j$ sends a random independent value $r_x$ that must have the same size as the key $k$~\cite[p.~1610]{kremer2002intensive}.
		Only in the last message, $o_j$ sends the actual decryption key $k$.
		Each message must be acknowledged by $c_i$ quicker than the decryption could be performed.
		After receiving $\text{ack}_k$ from $c_i$, $o_j$ ends the exchange.
		This leads to $c_i$ timing out, indicating that the exchange was completed.
		Thus, $c_i$ can proceed to compute the datum $d$ by decrypting $s$ with $k$.
		Lastly, the block is composed and published.
		Both parties store their respective non-repudiation evidence on their own machine.
	}
	\Description{A UML sequence diagram shows the steps of the protocol, as outlined in the caption.}
	\label{fig:protocol}
\end{figure}

The protocol is described in Fig.~\ref{fig:protocol}. For this scenario, we assume that $o_j$ returns the requested datum $d$ directly.
Alternatively, they could also return, e.g., the decryption key for a datum stored elsewhere.
To begin with, $o_j$ requests a one-time pseudonym from $c_i$. This is used to compose the block when the protocol completes. As the pseudonym is only relevant for $c_i$ to be able to identify the block, it does not need to be verified.
The number of steps $n$ is chosen at random by $o_j$. For less critical data, it can be reduced~\cite[p.~7]{markowitch1999probabilistic} to lower energy consumption or improve scalability.
Then, $o_j$ computes $n-1$ random independent values $r_x$ and a symmetric encryption key $k$.
Importantly, the random values must be of equal size to the chosen key.
Running $\text{enc}_k(d)$, they obtain the cipher $s$ that they send to $c_i$.
Now, in each step $x$, $o_j$ sends a message with one of the random values $r_x$ instead of the actual key $k$.
Only in the $n$\textsuperscript{th} and final step, $o_j$ sends $k$.~\cite{markowitch1999probabilistic, kremer2002intensive}
After completion of the exchange, $c_i$ can decrypt the cipher $s$ to obtain $d$.
Finally, $o_j$ composes a block for $u_{ij}(p(c_i) \to p(o_j))$ and publishes it.

Following \citeauthor{markowitch1999probabilistic}~\cite{markowitch1999probabilistic}, as $c_i$ cannot predict $n$, and if the chosen decryption function takes long enough to compute, they will not be able to get any meaningful data when cheating~\cite[p.~5]{markowitch1999probabilistic}.
Only after having received the last message, they can decrypt the cipher~\cite{shamir1979share, markowitch1999probabilistic}.
Now, each party holds non-repudiation evidence of the interaction. For $o_j$, this is \{~sign\textsubscript{c}(ack\textsubscript{$s$}, $o_j, t$), sign\textsubscript{c}(ack\textsubscript{$k$}, $o_j, t$)~\}, for $c_i$ it is \{~sign\textsubscript{o}($s, c_i, t$), sign\textsubscript{o}($k, c_i, t$)~\}~\cite[p.~1610]{kremer2002intensive}.

This protocol depends on the nodes being able to verify the authenticity of requests and, importantly, being protected against man-in-the-middle or eavesdropper attacks.
Therefore, each request is signed by the sender.
We do not aim to reinvent the wheel here, instead relying on the established HTTP over TLS standard~\cite{rfc2818httpovertls}.
This enables communication confidentiality and authenticity~\cite{krawczyk2013security}.
By utilizing the approach of a web of trust, as established in PGP~\cite{abdulrahman1997pgp}, nodes are fully independent of any trusted third party to verify certificates.
In that case, unknown certificates would be rejected and would need to be verified in-person.
Alternatively, if sensible for the specific deployment, a certificate authority can be used to sign the individual certificates used by each node to sign and encrypt its requests.
As these are often used in companies to enable the signing of internal emails or access to protected resources, no additional certification infrastructure is required in either case.

\subsubsection{Pseudonym Generation Algorithm}\label{sec:generation-algorithm}

The second core step towards our goal is the ability to generate unique one-time pseudonyms that guarantee unlinkability and enable proof of ownership without requiring a trusted third party.
Unlinkability is required for the data we store to be able to qualify as anonymous data (see Sec.~\ref{sec:pseudonymity-anonymity}).
Proof of ownership, on the other hand, enables the owners of the pseudonyms to exercise their rights as given by the GDPR~\cite[see][Art.~12.6]{eu2016gdpr}.

\citeauthor{florian2015sybil}~\cite{florian2015sybil} describe a pseudonym generation algorithm that serves as inspiration to our solution. Pseudonyms are guaranteed to be unlinkable to each other and to the real identity of the user. Furthermore, authenticity proofs enable proof of ownership, meaning that our requirements are met.
Beyond those properties, their algorithm provides sybil-resistance, which is achieved by requiring additional computational steps for joining a network and creating new pseudonyms~\cite[p.~68--69]{florian2015sybil}.
In our case, the additional property of sybil-resistance is not required, as there is no inherent danger in a user creating multiple pseudonyms (see Sec.~\ref{sec:generation-protocol}).
We therefore omit these additional complexities and simplify our algorithm accordingly. By that, we reduce its computational complexity and energy consumption to a minimum.

Therefore, we define our pseudonym generation algorithm as follows:
As part of the new-usage protocol (see Sec.~\ref{sec:generation-protocol}), the user has created a new RSA private-public key pair with a key size of 4096 bits. As discussed above, the chosen encryption method can be updated if a higher level of security is appropriate.
Now, to generate the one-time pseudonym, the collision-resistant and cryptographic hash function BLAKE2~\cite{aumasson2013blake2}, specifically BLAKE2s~\cite[p.~121]{aumasson2013blake2}, is applied to create a cryptographic message digest.
Concretely, the user hashes the public key of their key pair, with the resulting irreversible and cryptographically safe digest representing their one-time pseudonym.
BLAKE2s ensures a digest size of at most 32 bytes, which is important to minimize storage requirements.

The pseudonym generated with our algorithm then guarantees three important properties:
First, the owner of the pseudonym, and only the owner, can prove the authenticity of the pseudonym (shown below).
Second, the unique properties of the hash function guarantee uniqueness~\cite{applebaum2017low}.
Third, users only need to manage a single key pair for each block, significantly reducing the complexity of the operation and increasing its speed.

To enable proof of ownership, we make use of the asymmetric nature of the RSA key pair.
When a user wants to prove their ownership of a pseudonym, they sign a message with their private key and make available the corresponding public key.
The signed message proves that they are in possession of the private key.
The public key is then hashed by the recipient with the BLAKE2 hash function. If the result is the correct pseudonym, the ownership is proven.
As BLAKE2 is collision-resistant, it is infeasible for an attacker to guess a different string that would result in the same pseudonym.
Making matters even more secure, they would in addition need to crack the utilized RSA algorithm to be able to sign a message with a matching private key.

\subsubsection{Block Structure} \label{sec:concept-block-structure}

Next, we define the actual data stored in each block of the blockchain.
As we have stressed above, our goal is to not require any changes to the underlying blockchain software, thereby allowing our solution to be used with existing blockchains.

\begin{figure}[htbp]
	\centering
	\includegraphics[width=0.9\linewidth]{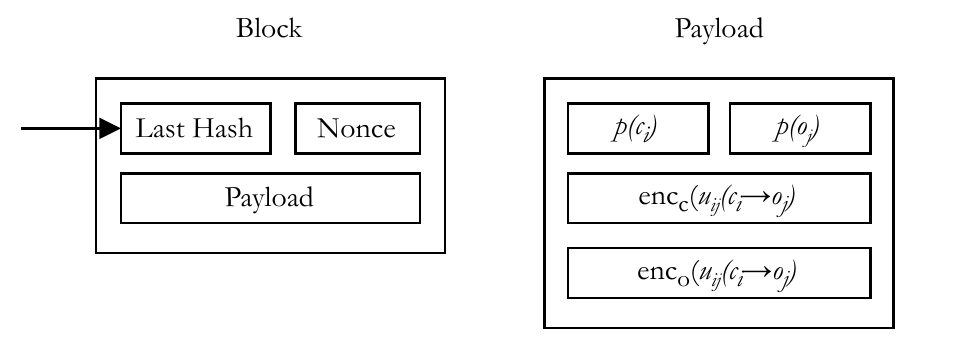}
	\caption{The components of each block (adapted from~\cite{nakamoto2008bitcoin}), for the exemplary usage $u_{ij}(c_i \to o_j)$. We replace the transaction history utilized in Bitcoin with a generic payload. The payload consists of the pseudonyms $p(c_i)$ and $p(o_j)$ of the data consumer and data owner, respectively. Furthermore, the logged usage is stored twice, once encrypted for the \emph{consumer} and once for the \emph{owner}.}
	\Description{Two boxes show the architecture of individual blocks in the blockchain as well as each payload. The block consists of the last hash, a nonce, and a payload. The payload consists of pseudonym(s) and encrypted data.}
	\label{fig:block}
\end{figure}

As an example, consider a new entry logging the usage $u_{ij}(c_i \to o_j)$.
Based on requirements \reqref{req:owners-verify} and \reqref{req:consumers-verify}, we need to store a one-time pseudonym for both the data consumer and data owner, to allow each party to efficiently query for entries concerning them.
In addition, both parties need to be able to access the stored usage information, while preventing third parties from reading it, following \reqref{req:confidentiality}.
As shown in Fig.~\ref{fig:block}, each block therefore contains a payload with the data consumer's pseudonym $p(c_i)$, the data owner's pseudonym $p(o_j)$, and two copies of the usage request.
This does not contain any identifiable information and only consists of the type of datum requested and a justification.
To provide confidentiality, each copy is encrypted with an encryption function $enc()$, once with the \emph{owner}'s and once with the \emph{consumer}'s one-time public key. Importantly, this key is not shared publicly and instead stored securely with the private key.
The chosen encryption should be regularly updated, but we require asymmetric (public-key) encryption~\cite{delfs2007public}.
As of this point, we recommend RSA~\cite{rivest1978method} with a key size of 4096 bits~\cite{lenstra2001selecting, kiviharju2017fog}.
Notably, though, each individual can choose their preferred key size, and therefore security level, themselves.

Besides allowing both parties to read the usage log, storing it twice is also important because the block is created by $o_j$ (see Sec.~\ref{sec:generation-protocol}). $c_i$ then needs to be able to verify the validity of the block. In case $o_j$ manipulates the stored entry, $c_i$ can utilize their copy of the usage and the non-repudiation evidence (see Sec.~\ref{sec:generation-protocol}) to defend themselves against the faked evidence.

\subsubsection{Deployment Model}
\label{sec:generation-deployment}

Finally, as we have hinted at above, the chosen deployment highly influences the privacy and security guarantees that can be given.
Our concept can be flexibly adapted and supports both centralized and peer-to-peer architectures.
As we aim to not be dependent on any trusted third party, though, our deployment architecture is fully decentralized.

\begin{figure}[htbp]
	\centering
	\includegraphics[width=0.75\linewidth]{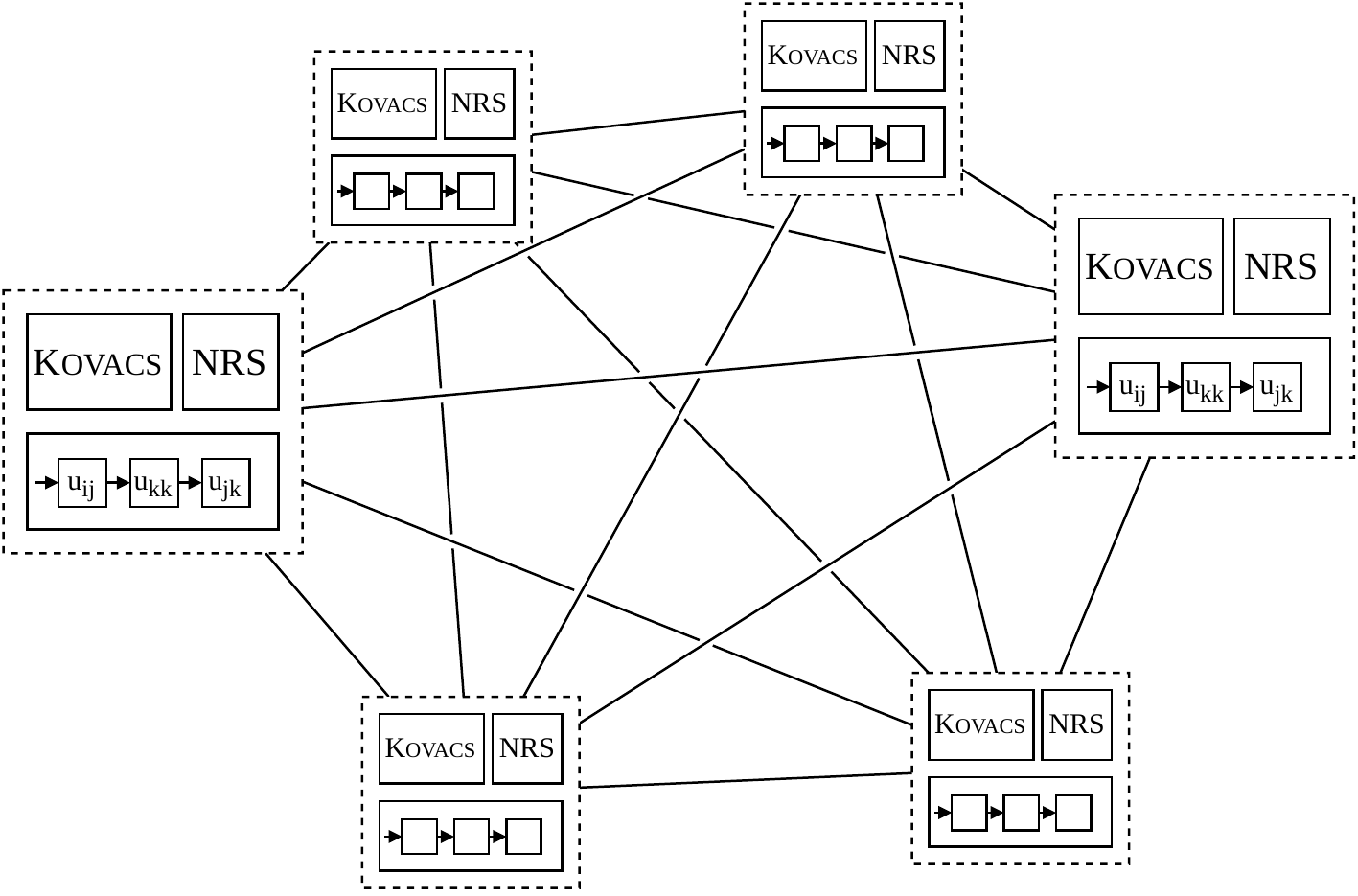}
	\caption{
		Deployment in a fully decentralized peer-to-peer architecture.
		Each node runs its own \Kovacs{} instance and private non-repudiation store NRS~\cite[see also][]{zhou1996observations}, while the blockchain copy is shared within the network~\cite[see also][]{nakamoto2008bitcoin}.
	}
	\Description{
		A network of Kovacs nodes is shown. They are all connected to each other, forming a star shape. Each node is symbolized by a dashed line and contains three boxes labeled Kovacs, NRS, and a third one containing a symbol representing blockchain.
	}
	\label{fig:deployment}
\end{figure}

The central component of the deployment is the \Kovacs{} system that handles data exchange, pseudonym provisioning, key management, and block creation.
Each node in the peer-to-peer network runs its own \Kovacs{} instance as well as a private non-repudiation store to store its RSA key pairs, pseudonyms, and non-repudiation evidences (see Secs.~\ref{sec:generation-protocol} and \ref{sec:generation-algorithm}).
The data exchange does not require an intermediary and utilizes the peer-to-peer network.
As the usage log blockchain is permissionless, it is shared between all nodes that want to participate in the network.
That means the architecture is fully decentralized, with each node communicating directly with other nodes both for data exchanges and blockchain updates (see Fig.~\ref{fig:deployment}).

The management of a large number of keys as required by our approach can itself become a privacy risk.
In theory, the nodes would not necessarily need to store their key pairs and used pseudonyms at all to reduce their attack surface.
Yet, these are important to enable requirements \reqref{req:owners-verify}, \reqref{req:consumers-verify}, and \reqref{req:gdpr}.
To be able to query for entries concerning their usages, users need to know which pseudonyms belong to them. In theory, they could iterate all blocks in the blockchain and simply try to decrypt their content, but this quickly becomes infeasible.
Additionally, to exercise their GDPR-awarded rights, users need to prove their ownership of a pseudonym, requiring them to be in possession of the respective private key for the specific block (see Sec.~\ref{sec:generation-algorithm}).
Otherwise, anyone could claim to be the owner of the encrypted data and, e.g., request its deletion.
Still, each user can choose on their own how to manage their keys. If they prefer the maximum level of security, they are free to, e.g., delete new keys immediately after usage.

	\section{Implementation}\label{sec:implementation}

Our implementation of \Kovacs{} represents one possible manifestation of our concept.
Some important facets have to be chosen for the concrete implementation and deployment scenario, namely the concrete blockchain, realizations of the underlying algorithms, as well as the integration with a complete network.
Along those facets, we therefore describe how we implement \Kovacs{} for our proof of concept and evaluation.
The source code of the final tool is available on GitHub under the MIT license.\footnote{\url{https://github.com/tum-i4/kovacs}}

\subsection{Blockchain}

First, the consensus mechanism and concrete blockchain for the implementation need to be chosen.
We aim for blockchain-agnosticism, therefore the concrete choices are considered an implementation detail.
We only outline our considerations and final choices below to ensure transparency and reproducibility regarding our evaluation.

\subsubsection{Consensus Mechanism}\label{sec:implementation:Blockchain:Consenus}

The choice of consensus mechanism was made between the commonly known proof of work (PoW)~\cite{nakamoto2008bitcoin}, proof of stake (PoS)~\cite{king2012ppcoin}, and proof of authority (PoA)~\cite{wood2015proof}. We arrive at the choice by process of elimination.
We cannot use PoA, as it would violate our requirements by requiring a central authority that creates and signs blocks~\cite{DeAngelis2018, loyoladaiqui2020proof}.
If we choose PoS, we need to have a currency that can be staked.
However, a usage log has no concept of ``currency''.
Accordingly, we use PoW as the consensus algorithm for our implementation.

\subsubsection{Blockchain Implementation}

We do not depend on smart contracts, which made our selection of a concrete blockchain more flexible.
Potential choices included Bitcoin~\cite{nakamoto2008bitcoin} and Ethereum~\cite{etherYellow}.
The widely used Hyperledger Fabric~\cite{androulaki2018hyperledger} was not considered, as it is based on PoA-based consensus.
Between the two, Ethereum offers multiple advantages for our use case,
namely a configurable hash difficulty
and support for the creation of private chains~\cite{ethereumPrivateChain}.
Therefore, we use it for our implementation.
As the client, we use the official implementation \emph{Go Ethereum} (Geth).

Note that, at the time of implementation, Ethereum still used PoW consensus.
Recently, Ethereum switched to PoS consensus~\cite{smith2023proof}.
When setting up a private network, it would therefore have to be configured to use PoW instead.
Alternatively, a different PoW-based blockchain can be used.

\subsection{Algorithms}

With the blockchain in place, we turn to the implementation choices for the core algorithms of the \Kovacs{} system.
Broadly, the algorithms are, of course, just implementations of our concept.
Yet, certain aspects may be realized in different ways depending on the application, which may impact, e.g., security or performance.
Therefore, in the following, we deliberate the concrete implementation choices for the identity verification, time-asymmetric encryption, block composition, and fake chatter algorithms.

\subsubsection{Identity Verification} \label{sec:implementation-identity-verification}

In order to enable the traceability of data accesses, which is fundamental for inverse transparency to enable accountability~\cite{zieglmeier2023itbd}, the data owner must know the data consumer's real identity.
This enables attribution of a data usage to the responsible party.
Thus, nodes need to be able to request and verify each other's identities.

We have noted in Sec.~\ref{sec:generation-protocol} that fully decentralized identity verification is possible by utilizing, e.g., a web of trust~\cite{abdulrahman1997pgp}.
Yet, in our work with industry partners, we found that most companies rely on institutional identity providers~(IdP) to realize single sign-on (SSO) for company-internal identity verification.
As the use case of inverse transparency is specifically tailored to the company-internal context, we therefore show how to integrate an institutional IdP for identity verification.
To minimize the risk of confidentiality attacks, we apply the concept of \emph{self-sovereign identity}~\cite[see, e.g.,][]{preukschat2021self, liu2020blockchain}.
That means that each party is issued a \emph{verifiable credential}~\cite{sporny2022verifiable} from the IdP, which it can present to exchange participants directly.

\begin{figure}[!ht]
	\begin{center}
		\centering
		\includegraphics[width=0.625\paperwidth]{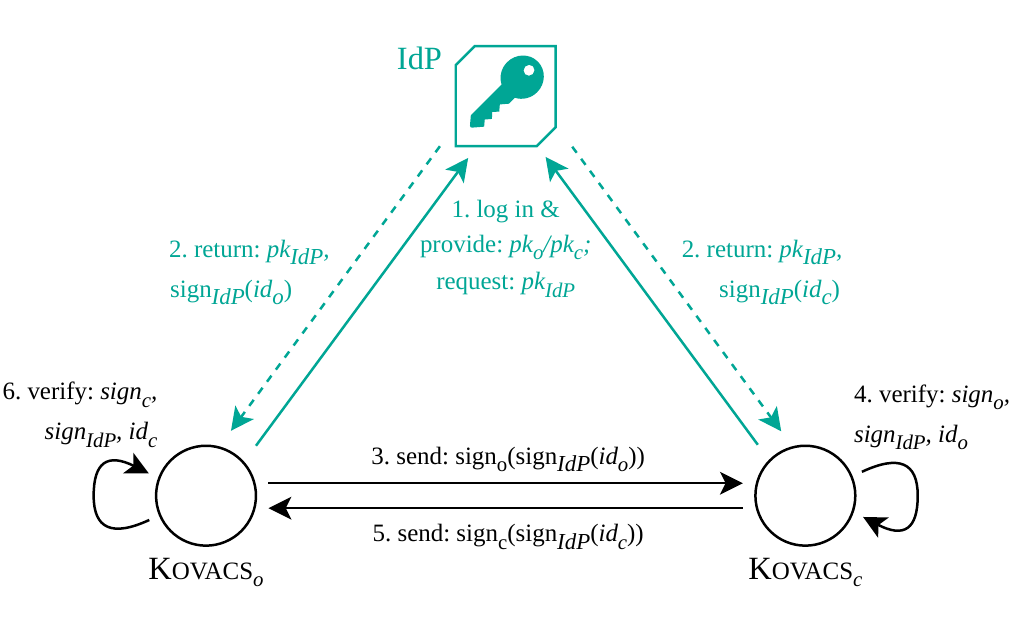}
		\caption{
			Identity verification steps for a data owner $o$ and a data consumer $c$, utilizing an external IdP to issue verifiable credentials~\cite[following][]{muhle2018survey};~\cite[see also][Chap.~3]{preukschat2021self}.
			The \Kovacs{} nodes of $o$ and $c$ log into the IdP once (1.) and send their public key to receive the corresponding verifiable credentials $id_o$ and $id_c$ as well as its public key $pk_{IdP}$ (2.).
			During future data exchanges, $o$ first provides their identity information to $c$ (3.), which $c$ verifies locally (4.).
			If the identity is confirmed, $c$ also provides their identity information (5.), which is verified as well by $o$ (6.).
		}
		\Description{
			A graph-like structure to visualize the identity verification steps.
			Three nodes interact with each other, visualized by arrows.
			The concrete steps of the algorithm that is visualized are described in this figure's caption.
		}
		\label{fig:implementation-identity-verification}
	\end{center}
\end{figure}

Enabling self-sovereign identity requires only small adaptations to the IdP and can otherwise utilize existing authentication infrastructure.
Following~\cite{muhle2018survey}, each party's \Kovacs{} node is issued a unique verifiable credential by the IdP on request.
To trigger this, it logs in and sends a public key to be associated with their identity.
The IdP then creates the verifiable credential containing of the user's IdP ID and their public key (the \emph{claims}), and its own signature verifying the authenticity of the credential (the \emph{proof}), and returns it to the requester~\cite[see also][Sec.~3.2]{sporny2022verifiable}.
Second, to prove their identity during the data exchange, the parties share their verifiable credential and sign them with the corresponding private key.
Hereby, the data owner has to prove their identity first, which protects the data consumer's privacy during peer search.
Each party can then compare the sender's signature to the public key on the credential and verify that the signature of the IdP on the credential is valid.
If both signatures are valid, that confirms that the identity used by the other party in the data exchange is authentic.
Following the principle of self-sovereign identity, this happens locally on each node, meaning the IdP cannot gain information on the data exchanges.

\subsubsection{Time-Asymmetric Encryption}\label{sec:implementation:non-repudiation}

After the identity verification, the actual data exchange occurs.
This exchange follows our new-usage protocol.
Critically, though, the fairness of the protocol depends on the decryption time of the transferred datum being longer than the chosen timeout duration~(see Sec.~\ref{sec:generation-protocol}).
We refer to encryption algorithms with this property as \emph{time-asymmetric}.
Such an algorithm would allow the data owner to simply encrypt the requested datum after they receive a data request.
Due to the longer decryption time, the timeout duration for data exchanges could be set just above the expected encryption time.
However, we were unable to find an encryption algorithm with this property in our research.
Thus, we have to assume that the encryption time is equal to the decryption time.

To our knowledge, there are two alternative options how a longer decryption time can still be realized, which have implications on \emph{when} the datum is encrypted and which \emph{timeout duration} should be chosen for the data exchange.
First, we can increase the timeout duration for the data exchange relative to the other steps of the protocol.
This solves the problem of symmetric encryption and decryption time, but results in a longer exchange duration.
The second variant is to (partially or completely) encrypt data \emph{before} the exchange begins, which tackles both problems.
Yet, assuming a reasonable number of options for which datum is actually requested, pre-encrypting all available data before the exchange is unrealistic.
Exacerbating this issue, the encryption key needs to be different for every transaction, which requires all data to be re-encrypted after each request.
Thus, a \emph{full} pre-encryption is infeasible.
Alternatively, we can move only \emph{parts of} the encryption routine before the start of an exchange.
These pre-computations must be independent of the specific requested datum to remove the need for re-encryption.
As this is the optimal solution for our scenario, we use this approach and implement a two-step encryption process, outlined below.

The encryption is split up into a time-consuming cipher key generation procedure and a fast en- and decryption.
Thus, the cipher keys can be precomputed, and only the encryption of the concrete datum has to be done at request time.
To realize a time-consuming key generation, we create a random string and hash it with a random salt using a password hashing algorithm.
These algorithms include key stretching functionality, which increases the time needed to calculate the hash~\cite{Kelsey1998KeyStreching}.
The hash resulting from this operation is the cipher key.
Importantly, the data owner only sends the random string and salt, requiring the data consumer to repeat the time-consuming key generation before being able to decrypt the datum.
For the implementation of our proposed encryption algorithm, we use bcrypt~\cite{Provos1999Bcrypt} as the password hashing algorithm and AES-256 GCM~\cite{NIST2001AES} for the symmetric encryption.
bcrypt and AES are widely adopted and their security guarantees have been verified on multiple occasions~\cite[see, e.g.,][]{Batubara2021BcryptSecurity, Singh2013AESSecurity}.
Our choice of AES GCM specifically is based on its guarantees regarding the integrity and confidentiality of data~\cite[p.~1]{Dworkin2007GCM}.

\subsubsection{Block Composition}\label{sec:implementation:Blockchain:export}

The final step of the new-usage protocol is the block composition (see Sec.~\ref{sec:generation-protocol}).
As described in Sec.~\ref{sec:concept-block-structure}, we store the usage logs in transactions that are added in blocks to the blockchain.
For simplicity, we do not rearchitect the blocks and just use regular exchange transactions to store the usage logs.
This allows our system to even be used with an arbitrary public blockchain as a storage backend for increased security.
Our choice requires us to mine two blocks for each logged usage: one to earn ``currency'' and a second to publish the usage log transaction.
We require currency to be able to conduct a transaction that contains the usage log.
Due to our unlinkability requirement, we create a new account for this purpose for each new usage log.
Concretely, that means the following steps are conducted when a block is composed.

First, a temporary account is created and registered to receive mining rewards.
Then, a block is mined to earn ``currency''.
Now, a transaction is added that sends the generated reward from the temporary account to a hardcoded address.
This transaction contains the newly created usage log.
Therefore, it is not important who the receiver of the transaction is, since we only consider the metadata.
The usage log transaction is still pending, meaning it is not yet stored in the blockchain.
Thus, a second block is mined which contains the transaction.
Finally, the temporary account is deleted.

\subsubsection{Fake Chatter} \label{sec:implementation-fake-chatter}

Even though all communications are encrypted, an eavesdropping attacker could relatively easily trace newly published usage logs on the blockchain to participating nodes if only few exchanges take place.
This attribution is possible because if exactly one block is published just after an exchange has ended, this block refers almost certainly to said exchange.
Such an attribution would weaken the unlinkability of logs, though.

To solve this issue, we implement the optional \emph{fake chatter} protocol.
That means that nodes can complete fabricated ``exchanges'' that generate traffic, but do not add new blocks to the blockchain.
Specifically, the data consumer creates a peer that completes an exchange as usual and additional peers that imitate data exchanges with random data owners.
During such an exchange, the data consumer informs the data owner that this is not a real data exchange.
Accordingly, no actual data are shared and no usage log is created.
That also means that the blockchain is not written to, ensuring that it is not congested.
Since the communication between the peers is encrypted, an attacker cannot distinguish between fake chatter and real exchanges.
Hence, they would be unable to attribute a usage log to an exchange.
If a sufficient number of concurrent exchanges take place, the protocol can be deactivated.
Consequently, the peers' privacy is protected in both cases.

\subsection{Integration}

As the final aspect of our implementation, we need to consider the integration of the \Kovacs{} system with a complete network.
In the following, we therefore describe how the peer-to-peer network of nodes is created without a trusted third party and explain how we adapted the open-source \emph{Revolori} SSO server to enable our identity verification algorithm.

\subsubsection{Peer Discovery}\label{sec:implementation:P2P}

We use \emph{libp2p} as our peer-to-peer library since it offers encrypted communication and peer discovery for both structured and unstructured networks.
For a structured network, libp2p requires a bootstrap node~\cite{libp2pDHT_BOOTSTRAP}.
Thus, we do not use this approach since it would violate our goal of decentralization.
As an alternative, libp2p also offers peer discovery in an unstructured network implemented with a flooding algorithm, which does not rely on any centralized service~\cite{libp2pMDNS}.
While flooding suffers from longer search times and likely worse scalability, we prioritize decentralization over performance.
Accordingly, we implement an unstructured network.

\subsubsection{Identity Verification Server}

Finally, we adapt the existing \emph{inverse transparency toolchain}~\cite{zieglmeier2021trustworthy, zieglmeier2023toolchain} to integrate the \Kovacs{} system with a realistic identity verification server.
In our realization of inverse transparency, we completely redesign the original fully centralized architecture to be as decentralized as reasonable.
As the only centralized component, we use the SSO server \emph{Revolori} as a stand-in for a company-internal IdP.
We adapted it for our use case, adding functionality for the creation of verifiable credentials.
Furthermore, we added an API endpoint that allows access to its public key, enabling local identity verification (see Sec.~\ref{sec:implementation-identity-verification}).
These changes are non-invasive to the functionality of \emph{Revolori}.
Thereby, we show how \Kovacs{} can be integrated with the existing inverse transparency toolchain.
Our adaptations to \emph{Revolori} are available on GitHub.\footnote{\url{https://github.com/tum-i4/inverse-transparency/tree/kovacs}}

\subsection{System Model}

\begin{figure}[!ht]
	\begin{center}
		\centering
		\includegraphics[width=\textwidth]{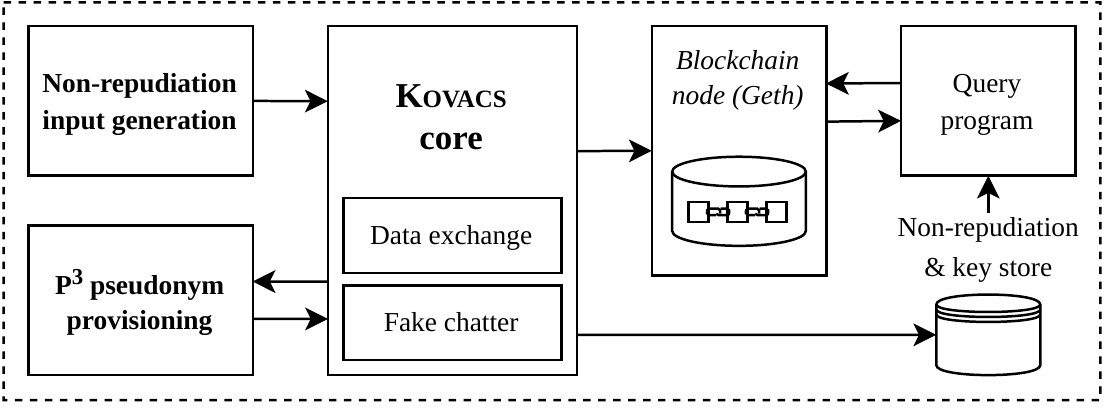}
		\caption{
			A \Kovacs{} node.
			The arrows represent communication and data flow between the components.
			Bold names denote the three logic components: \Kovacs{} core, non-repudiation input generation, and \PPP{} pseudonym provisioning.
			The name of the Geth node is italicized to denote it is a third party tool.
		}
		\Description{
			A box with dashed outline is shown that contains other boxes with regular outlines, representing the Kovacs components.
			They are connected with arrows.
			In the middle, the Kovacs core component contains two sub-components: data exchange and fake chatter.
			It receives data from the components non-repudiation input generation and P3 pseudonym provisioning to the left of it.
			The core sends data to the non-repudiation and key store as well as the blockchain node (Geth), the latter again italicized (third party).
			Finally, the query program located top right receives data from the store and interacts with the blockchain node.
		}
		\label{fig:implementation-kovacs-architecture}
	\end{center}
\end{figure}

Finally, we connect these components into a complete system.
Fig.~\ref{fig:implementation-kovacs-architecture} is a model of the \Kovacs{} system, outlining its components and their interactions.
As the main component, the \emph{\Kovacs{} core} handles interaction within the node and with the peer-to-peer network (see Sec.~\ref{sec:generation-deployment}).
For data exchanges it first requests input data from the \emph{non-repudiation input generation} component, which generates the parameters for the new-usage protocol (see Sec.~\ref{sec:generation-protocol}).
Furthermore, it requests one-time pseudonyms from the \emph{\PPP{} pseudonym provisioning} component (see Sec.~\ref{sec:generation-algorithm}).
The \emph{query program}, finally, is utilized by the user to access the stored usage log.
To do so, it loads all used pseudonyms and the decryption keys for the stored data from the \emph{non-repudiation \& key store}.
Then, it queries the blockchain via the \emph{blockchain node} to retrieve the requested usage log entries.

	\section{Evaluation}

To evaluate the \Kovacs{} system, we first analyze the security of the concept and any implementations based on it.
Second, we assess the GDPR compliance, specifically focusing on the data stored in the blockchain.
Third, we benchmark the performance of our implementation for the most time critical operations.
Finally, we measure its scalability for increasing numbers of log entries.

\subsection{Security Analysis} \label{sec:evaluation-security}

We begin by analyzing the security of our system based on two core aspects: its robustness against attacks and the protocol confidentiality.

\subsubsection{Robustness Against Attacks}\label{sec:analysis-attacks}

We have described an adversarial model in Sec.~\ref{sec:adversarial-model}, with adversary $\alpha$ trying to subvert the integrity and confidentiality of the stored logs.
Specifically, they try to conduct six attacks.
For each, we analyze the robustness of our approach against the attack and potential implications.

First, $\alpha$ tries to (a) repudiate a logged usage $u_{ij}(c_i \to o_j)$ after receiving the datum.
To prevent this, we need to guarantee \emph{non-repudiation of receipt}~\cite{zhou1996observations}.
Our new-usage protocol is built on the \citeauthor{markowitch1999probabilistic} non-repudation protocol, inheriting its security properties.
As shown by \citeauthor{markowitch1999probabilistic}~\cite{markowitch1999probabilistic} and later confirmed by \citeauthor{aldini2002security}~\cite{aldini2002security}, the protocol can guarantee non-repudiation of receipt based on the chosen success parameter $\theta$, which influences the choice of the number of rounds $n$.
As $\theta$ is chosen by the data owner, they can configure the protocol to make it infeasible for $\alpha$ to repudiate a logged usage, preventing $\alpha$ to repudiate the usage.
We discuss the performance implications of this in Sec.~\ref{sec:evaluation:perf:exchangeDuration}.

Second, $\alpha$ tries to (b) fabricate an entry $u_{ij}(c_i \to o_j)$ for a usage that did not occur, incriminating $c_i$.
This requires them to successfully fabricate a corresponding \emph{non-repudiation of origin} evidence.
In our case, this is \{~sign\textsubscript{c}(ack\textsubscript{$s$}, $o_j, t$), sign\textsubscript{c}(ack\textsubscript{$k$}, $o_j, t$)~\} (see Sec.~\ref{sec:generation-protocol}).
For this attack to be feasible, let us assume that $\alpha$ previously exchanged data with $c_i$, thereby receiving their verifiable credential.
Given this, fabricating the non-repudiation of origin proof would still necessitate $\alpha$ to retroactively calculate the RSA private key of $c_i$ matching the given verifiable credential.
The private key is required to generate the proofs described above.
The security of the utilized RSA has been shown previously~\cite[see, e.g.,][]{mahto2016security}.
Furthermore, our selected key size of 4096 bits provides higher-than-usual security~\cite{kiviharju2017fog}.
This makes it clearly infeasible for $\alpha$ to compute the RSA private key, even if they can utilize significant computing power.
Should the chosen key size should become insufficient with rising computing power, though, it can be flexibly increased to harden the security.

Third, $\alpha$ tries to (c) derive from any usage $u_{ij}(c_i \to o_j)$ the identity of $c_i$ or $o_j$.
As the transaction pseudonyms for both $c_i$ and $o_j$ are created with the same algorithm, this attack depends on being able to reverse the employed pseudonym generation. The one-time cryptographic security of the utilized BLAKE2 algorithm has been shown~\cite[see, e.g.,][]{aumasson2013blake2, luykx2016security}, guaranteeing it to be irreversible.
Furthermore, each transaction uses a new key pair and pseudonym, so their unlinkability is ensured even if the pseudonyms were reversible.
This means that multiple usage logs cannot be linked.
The metadata of an individual usage only contains keys that are not connected to the identity of the user, though.
Therefore, $\alpha$ can gain no information on the identity of $c_i$ or $o_j$ from it.

Fourth, $\alpha$ tries to (d) associate any two usages $u_{ij}(c_i \to o_j)$, $u_{ik}(c_i \to o_k)$ with each other, revealing their association with a single \emph{consumer} and, fifth, (e) associate any two usages $u_{ji}(c_j \to o_i)$, $u_{ki}(c_k \to o_i)$ with each other, thereby leaking their association with a single \emph{owner}.
We can discuss both attacks together, as they hinge on the same security mechanism.
The cryptographic security of our algorithm is guaranteed by the cryptographic security of the two underlying algorithms RSA and BLAKE2, which has been shown for both~\cite[see][]{mahto2016security, luykx2016security}.
Therefore, this again depends on the ability of $\alpha$ to reverse the transaction pseudonym generation. As we have shown above, this can be considered infeasible.

Finally, $\alpha$ tries to (f) leak the identity of $c_j$ for a stored usage $u_{jj}(p(c_j) \to p(o_j))$ with $\alpha = o_j$, after $c_j$ has exercised their right to erasure.
In the last section, we have detailed that the association of $c_j$ to their pseudonym is known to $o_j$ for blocks storing usages of data that $o_j$ owns.
We have no way to technically force $o_j$ to delete this association when $c_j$ exercises their right to erasure.
Now, let us assume that $o_j$ does not delete this data and wants to utilize their knowledge, e.g., by publishing the real identity of $c_j$ and their one-time pseudonym $p(c_j)$. By itself, this proves nothing, as there is no technical relationship between the pseudonym and the identity of $c_j$~(see Sec.~\ref{sec:generation-algorithm}).
To actually prove the association of $c_j$ to $u_{jj}$, $o_j$ therefore needs to publish their non-repudiation evidence~(see Sec.~\ref{sec:generation-protocol}).
This evidence, by design, contains their own identity (through their signature) as well~\cite[pp.~5--6]{markowitch1999probabilistic}.
This means that $o_j$ would automatically also leak their own identity, making them legally liable.
As this scenario is covered by legislation and can be prosecuted accordingly, we consider it a non-issue for most cases.
Still, for the most secretive of environments, this might not be enough of a guarantee.

\subsubsection{Protocol Confidentiality}

For the highest-security deployments, our peer-to-peer architecture enables pseudonym generation and block creation without necessitating a trusted third party, mitigating most attack vectors on the integrity and confidentiality of data.
Two potential attack vectors on the confidentiality of the exchanged information remain: The messages sent on block creation, and the block update.

Firstly, when creating a new block for a usage $u_{ij}(c_i \to o_j)$ following the protocol (see Sec.~\ref{sec:generation-protocol}), communication between $c_i$ and $o_j$ has to occur. Even though HTTP over TLS is utilized, which prevents $\alpha$ from listening in as an eavesdropper~\cite{krawczyk2013security}, they may still deduce that there is some usage association between the nodes.
To address this, we have implemented the \emph{fake chatter} protocol (see Sec.~\ref{sec:implementation-fake-chatter}).
This works much the same way as the regular new-usage protocol, only that a special non-existent datum $d_0$ is requested.
Then, both parties understand that this is just a fake request, and no actual block is added to the blockchain.
This fake protocol can be run by nodes in randomized intervals, choosing arbitrary other nodes to request $d_0$ from.
Thereby, we hide real requests in the noise of these fake requests.

What remains then is the block creation. Even if fake protocols are run regularly, $\alpha$ could simply watch for blockchain updates and derive from those which two nodes were responsible for the new block.
This is possible because the block is added right after the protocol has concluded.
The simplest mitigation of this is to add a random wait before the block is added.
Then, plausible deniability is enabled, as there are a sufficient number of other potential users that might have been responsible. In fact, nodes might even wait for a certain number of block updates before publishing their update. Here, too, each node can decide itself the level of confidentiality it requires, and act accordingly.
Furthermore, traditional blockchain algorithms already (indirectly) protect from $\alpha$ understanding the originator of a blockchain update. As the architecture is designed to be peer-to-peer, the mere fact that a node sends a block update does not give $\alpha$ any additional information about its creation.
Nodes forward block updates to other nodes, so the specific node that sends $\alpha$ the update may also simply have forwarded it~\cite{nakamoto2008bitcoin, zheng2017overview}.

\subsubsection{Conclusion}

The \Kovacs{} system is robust against the most likely attacks as defined in our adversarial model.
Furthermore, the new-usage protocol can easily be adapted to fulfill even the highest requirements towards information security.
We conclude that usage logs created by \Kovacs{} provide sufficient security even for highly adversarial deployments.

\subsection{GDPR Compliance} \label{sec:evaluation-gdpr}

In the following, we analyze the data stored in the blockchain, assessing the compliance of our solution with the GDPR.

\subsubsection{Prerequisites}

We have discussed above (see Sec.~\ref{sec:pseudonymity-anonymity}) that data can be considered pseudonymous and anonymous. When a possibility for re-identification exists, they count as pseudonymous and therefore fall under the provisions of the GDPR~\cite{enisa2019recommendations}.
To comply with the GDPR, we need to enable data subjects to exercise their GDPR rights.
We also found, in line with legal analyses, that the main GDPR right that is technically challenging when utilizing blockchain is the \emph{right to erasure}~\cite{pagallo2018chronicle, tatar2020law} (see Sec.~\ref{sec:concept-requirements}).
Accordingly, we analyze conformance with the GDPR's right to erasure in the following.

\subsubsection{Analysis}

Each block in our approach gets its own transaction pseudonym. We have shown in Sec.~\ref{sec:generation-algorithm} that these pseudonyms are unlinkable to the individual's identity and to each other.
For a usage $u_{ij}(p_x(c_i) \to p_y(o_j))$, only $c_i$ and $o_j$ know the association of the other party's single one-time pseudonym ($p_x$ or $p_y$) to their real-world identity.
In fact, $c_i$ and $o_j$ have to prove their identity to each other in the first step of the new-usage protocol (see Sec.~\ref{sec:generation-protocol}).
Due to the guaranteed unlinkability, this link is only given for the single pseudonym created for that block, i.e. $o_j$ only knows the link $p_x(c_i) \leftrightarrow c_i$.
This case is covered by the GDPR provisions as there is an identifiable data controller~\cite[c.f.][]{vaneecke2018blockchain}.
The association is not stored in the blockchain, but only on the nodes of $c_i$ and $o_j$. That means the request for deletion simply has to be forwarded to them.
Considering an adversarial user, they might just not fulfill that request for deletion (see also Sec.~\ref{sec:analysis-attacks}).
At first glance, this could imply that our protocol does not offer an advantage over modifying the blockchain and asking all nodes to delete their old copy.
Importantly, though, we do not deal with \emph{unknown} nodes.
When a deletion request is raised, the data subject \emph{knows} the identity of the offending user, and can \emph{prove} it (see Sec.~\ref{sec:generation-protocol}).
That means, the individual can then be made responsible for deletion under the GDPR, and can be sued in case they do not follow through.

As soon as the association of the individual's identity to the pseudonym has been deleted, the data stored in the blockchain are anonymized.
Then, they do not qualify as personal data anymore (see also \cite{enisa2019recommendations} and Sec.~\ref{sec:pseudonymity-anonymity}), satisfying the right to erasure~\cite[Rec.~26]{eu2016gdpr}.

\subsubsection{Conclusion}

As we could show, our solution satisfies the GDPR's right to erasure.
Thereby, it solves the central challenge for blockchain arising from the GDPR~\cite{pagallo2018chronicle}.
Indirectly, this also enables the right to rectification, by deleting an entry and adding the rectified version.
The other GDPR rights are either not applicable in our case or trivially enabled from a technical point of view (e.g., right of access; see Sec.~\ref{sec:concept-requirements}).
We conclude that the \Kovacs{} system fulfills the technical requirements for GDPR compliance.

\subsection{Performance} \label{sec:evaluation:performance}

After the theoretical analysis, we assess the performance of our \Kovacs{} implementation.
All measurements were taken on an eight-core Ryzen 7 5800X with 16GB of DDR4 RAM running at 3600 MHz.
The timeout duration of the non-repudiation protocol was set to three seconds. Thus, the decryption time needed to be at least three seconds, which was achieved with a bcrypt hash difficulty of 16.
All nodes were run in Docker containers. Our adapted \emph{Revolori} SSO was run on the same system, with nginx acting as a reverse proxy to allow the nodes to connect to it using the host system's network address.
Geth was configured to be a full node, meaning that each node has a copy of the entire blockchain.
Unless otherwise stated, the network was run with fifty peers.

We evaluate two concrete scenarios, namely the exchange duration, which differs for data owner and data consumer, as well as the log entry append time, which is only relevant for the data owner.

\subsubsection{Exchange Duration}\label{sec:evaluation:perf:exchangeDuration}

To analyze the exchange duration, we first need to define its beginning and end. From the perspective of a data consumer, the exchange begins with the start of their node and ends with them receiving and decrypting the requested datum.
From the perspective of a data owner, the exchange begins when a data consumer connects to their node and ends after the new-usage protocol is completed and the usage log entry is appended to the blockchain.

\begin{figure}[htbp]
	\centering
	\includegraphics[width=\textwidth]{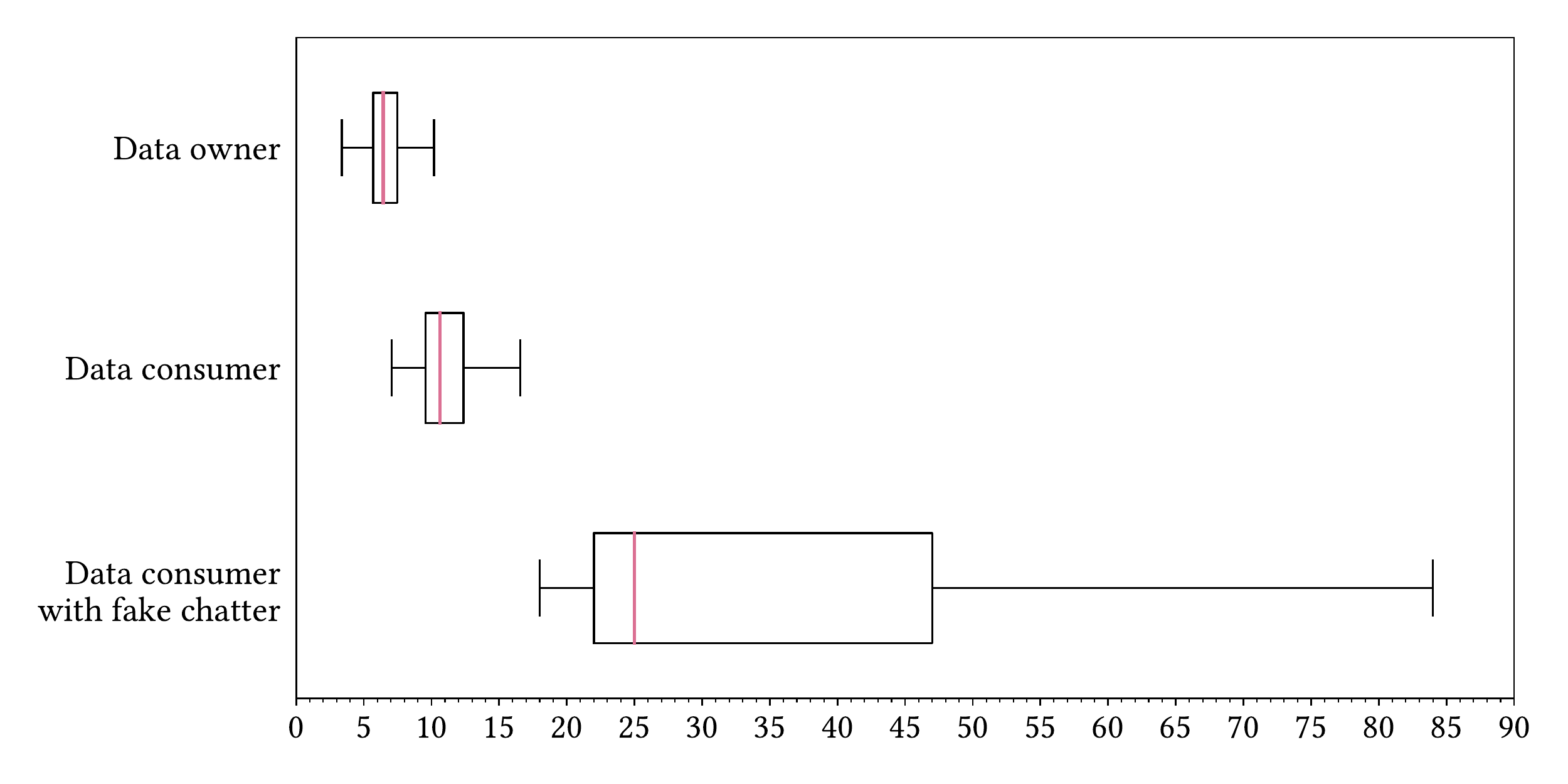}
	\caption{
		Duration (in seconds) of an exchange from the perspective of the data owner, the data consumer, and the data consumer with fake chatter enabled (each measured over 1500 runs).
		The medians of approximately 6.4 seconds for the data owner, 10.6 seconds for the data consumer, and 25.0 seconds for the data consumer with fake chatter are marked with pink lines.
		The whiskers extend to $1.5 \times IQR$, meaning they show the 25\textsuperscript{th} and 75\textsuperscript{th} percentiles.
		Outliers beyond the whiskers are not shown.
	}
	\Description{
		Box-and-whiskers plot of the exchange durations from the perspective of the data owner, data consumer, and data consumer with fake chatter enabled.
		The plot shows the time in seconds.
		For the data owner, the lower whisker sits at around 3.5 seconds, the upper whisker at around 10.5 seconds. The box for the data owner spans from around 5.8 to around 7.6 seconds.
		For the data consumer, the lower whisker sits at around 7.2 seconds, the upper whisker at around 16.7 seconds. The box for the data consumer spans from around 9.6 to around 12.5 seconds.
		For the data consumer with fake chatter, the lower whisker sits at around 18 seconds, the upper whisker at around 84 seconds. The box for the data consumer spans from around 22 to around 47 seconds.
	}
	\label{fig:evaluation:performance-exchangeDuration}
\end{figure}

Measured over 1500 runs, we find that the median exchange durations for data owner and consumer are 6.4 and 10.6 seconds, respectively (see Fig.~\ref{fig:evaluation:performance-exchangeDuration}), averaging 6.6 and 12.2 seconds.
Thus, the exchange is approximately 5.6 seconds longer for the data consumer.
This additional time for the data consumer is mainly spent waiting to time out and decrypting the received datum after the exchange has ended.
To contextualize these results, we calculate the additional time effort for our approach compared to simple, repudiable peer-to-peer data sharing.
For that, we measure the time for the additional required steps of our exchange, namely setup steps and peer search ($\approx 3.1$~s, only data consumer), the new-usage protocol ($\approx 4.2$~s for data owner, $\approx 9.0$~s for consumer), and account and mining operations relating to the blockchain ($\approx 2.5$~s, only data owner).
The identity verification is negligible with $< 0.5$ milliseconds.
This shows that the largest proportion of time is spent on the new-usage protocol. Depending on network latency, this can rise further.

As described in Sec.~\ref{sec:implementation-fake-chatter}, we implement the optional fake chatter protocol to increase the confidentiality of interactions.
Naturally, it also reduces the performance of data exchanges, though.
In our implementation, only the data consumer performs fake chatter, meaning the exchange duration is only affected for them.
Figure~\ref{fig:evaluation:performance-exchangeDuration} accordingly also shows the exchange duration from the perspective of the data consumer with fake chatter enabled.
We aim to give an upper bound of the required time and therefore allowed fake chatter to trigger between 12 and up to 321 additional fake connections per exchange.
This number is intentionally very high and can be lowered in practice.
With this, we find that an exchange with fake chatter takes approximately $2.4 \times$ longer for the data consumer than without it, resulting in a median exchange duration of 25.0 seconds.

\subsubsection{Appending a Log Entry}\label{sec:evaluation:perf:export}

\begin{figure}[htbp]
	\centering
	\includegraphics[width=\linewidth]{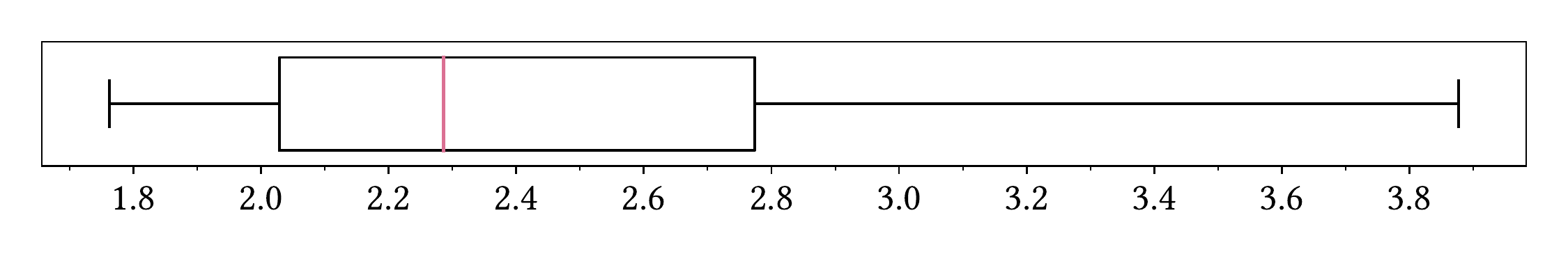}
	\caption{
		Duration (in seconds) of a blockchain append, measured over 1500 runs.
		The median of approximately 2.3 s is marked with a pink line.
		The whiskers extend to $1.5 \times IQR$, meaning they show the 25\textsuperscript{th} and 75\textsuperscript{th} percentiles.
		Outliers beyond the whiskers are not shown.
	}
	\Description{
		Box-and-whiskers plot of the blockchain export durations measured over 1500 runs.
		The plot shows the time in seconds. The lower whisker sits at 1.76 seconds, the upper whisker at around 3.9 seconds. The box spans from about 2.05 to 2.78 seconds.
	}
	\label{fig:evaluation:performance-write}
\end{figure}

After the exchange is completed, the data owner appends the newly created usage log entry to the blockchain.
Measured over 1500 runs, we find that the median blockchain append takes 2.3 seconds (see Fig.~\ref{fig:evaluation:performance-write}).
In some cases, we measured durations of up to 11.9 seconds, but these are rare.
We also find that the first blockchain append is consistently significantly slower than the median.
We did not investigate further, as we attribute this slowdown to initiation overhead of the Geth client.

To understand which individual operations take up most time, we can roughly split the blockchain append into three steps:
(1) Mining, to earn currency and store the transaction containing the usage log,
(2) creating and unlocking an account to create a transaction, and
(3) creating the transaction that will store the usage log.
Since we use a PoW consensus algorithm, one could expect mining to be the main reason for the slow blockchain append.
However, our benchmarks reveal that mining accounts for less than half of the time spent (on average $1.01$~s). The remaining time is spent creating and unlocking the account (on average $1.47$~s), which is performed by Geth.
Finally, the time spent creating a transaction is negligible at about $2$ milliseconds.

\subsubsection{Conclusion}

Both the exchange duration and blockchain append times of \Kovacs{} are notably slow, being measured in seconds.
While this is expected, as we focus on increased security, we need to note that this is a clear trade-off.
For scenarios with lower security requirements, the significantly increased exchange duration specifically can be disqualifying.
Yet, given our primary goals of decentralization and high security, we consider the time taken to still be reasonable.
Furthermore, this time does not increase even if data from multiple data owners need to be requested, as requests can be parallelized.
That means that these operations do not impact scalability.

\subsection{Scalability} \label{sec:evaluation:scalability}

Finally, we evaluate the scalability of our \Kovacs{} implementation.
As above, all measurements were taken on an eight-core Ryzen 7 5800X with 16GB of DDR4 RAM running at 3600 MHz, and all nodes were run in Docker containers.
Concretely, we evaluate the log retrieval time as well as the storage requirements for increasing numbers of log entries.

\subsubsection{Retrieving Usage Logs}\label{sec:evaluation:perf:query}

One of the advantages of \Kovacs{} compared to other blockchain-based secure usage logs is that searching through log entries does not require decrypting each block.
Therefore, one would expect good scaling behavior when retrieving usage logs.
To evaluate this, we measured both retrieval of a single, random usage log, and retrieval of all usage logs.
We ran each step 50 times and averaged the results.

\begin{figure}[htbp]
	\centering
	\begin{subfigure}[b]{0.49\textwidth}
		\centering
		\includegraphics[width=\textwidth]{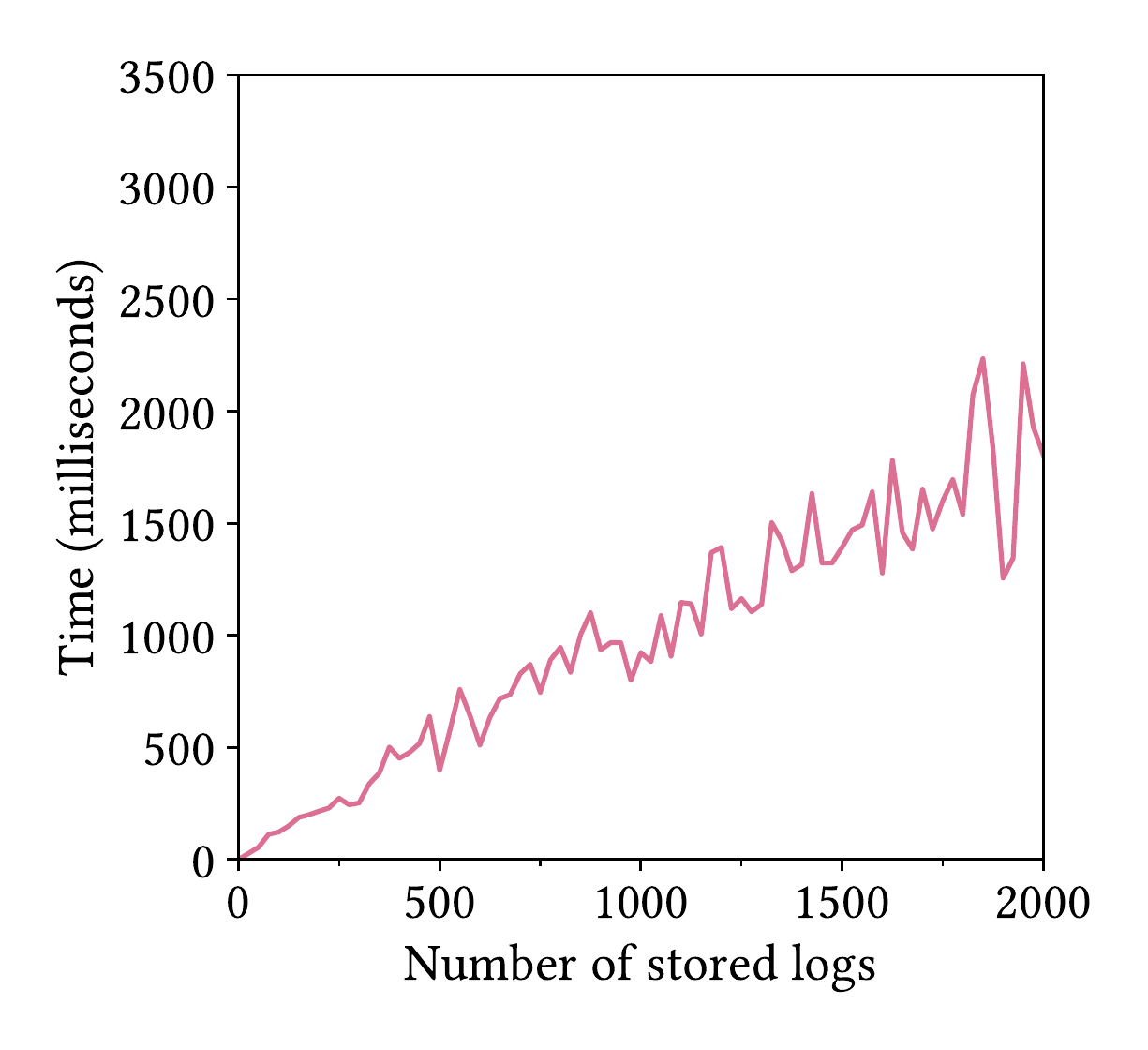}
		\caption{
			Time to retrieve a random usage log.
		}
		\Description{
			Line graph of the search time for a random usage log.
			The line for Kovacs rises from 0 up to about 1.9 seconds for 2000 logs relatively linearly. It does show strong variations, though.
		}
		\label{fig:evaluation:performance-querySingle}
	\end{subfigure}
	\hfill
	\begin{subfigure}[b]{0.49\textwidth}
		\centering
		\includegraphics[width=\textwidth]{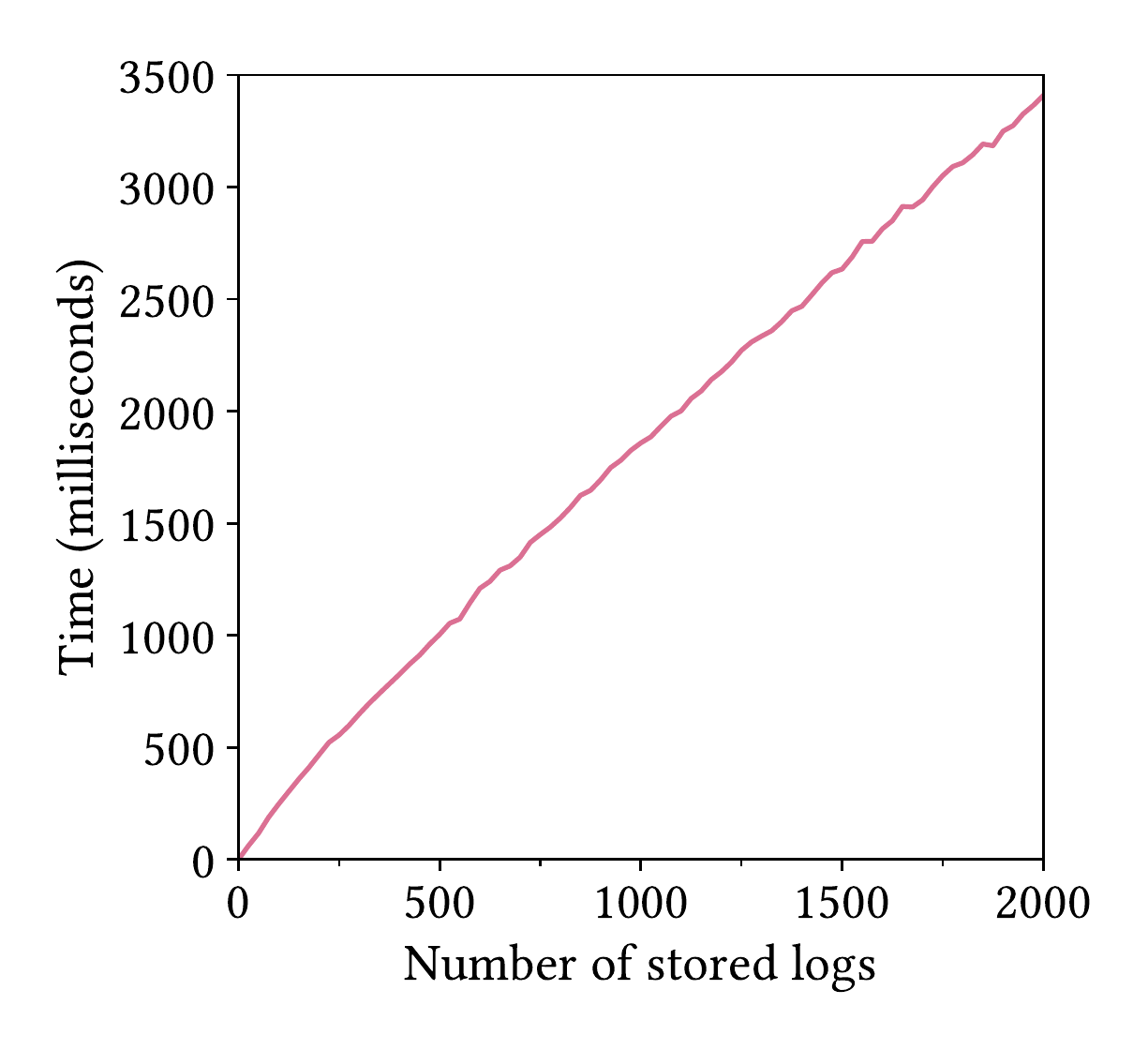}
		\caption{
			Time to retrieve all usage logs.
		}
		\Description{
			Line graph of the search time for all usage logs.
			The line for Kovacs rises from 0 up to about 3.4 seconds for 2000 logs almost linearly.
		}
		\label{fig:evaluation:performance-queryAll}
	\end{subfigure}
	\caption{
		Average retrieval time (in milliseconds) of a specific but random usage log entry (left) and all usage log entries (right), measured for log sizes of 25 through 2000 stored logs, with a step size of 25, and with 50 repetitions for each step.
	}
	\Description{This figure just groups the two graphs for the search time side by side.}
	\label{fig:evaluation:queryCombined}
\end{figure}

Overall, we find that \Kovacs{} returns the result in hundreds of milliseconds in both cases, with, e.g., retrieval of one log out of 1000 stored logs taking on average $0.97$ seconds (see Fig.~\ref{fig:evaluation:performance-querySingle}), with a median of $0.92$ seconds.
Retrieving all logs in this case, as a comparison, takes on average $1.88$ seconds (see Fig.~\ref{fig:evaluation:performance-queryAll}), with a median of $1.86$ seconds.
The high variance that can be observed for the single log retrieval task (see Fig.~\ref{fig:evaluation:performance-querySingle}) can be explained with the random log selection.
The newer the selected log is, the more blocks may need to be searched before finding a match, which directly influences the retrieval time.
More interesting than the absolute number is that the retrieval time in \Kovacs{} increases only linearly with the number of stored logs.
We can estimate the scaling behavior with linear regression.
For retrieval of a single, random entry, we obtain a slope of approximately $0.91 \times x$~ms with $R^2 > 0.92$.
For retrieval of all entries, we obtain $\approx 1.66 \times x$~ms with $R^2 > 0.99$.

Compared to a centralized database that is stored on a remote machine, we expect the overall retrieval time to be relatively competitive in real-world scenarios, as \Kovacs{} does not need to retrieve data over the network.
Instead, it can directly query the local blockchain copy.
This, combined with its efficiently queryable blockchain, positively impacts the scalability of \Kovacs{}.

\subsubsection{Storage Requirements}\label{sec:evaluation:perf:storage}

Finally, we consider the storage requirements of our \Kovacs{} implementation.
This is more relevant in our case than with traditional centralized data stores, as each node stores a copy of the full blockchain.
Therefore, we measured the total log size and calculated the average size per log as well.

\begin{figure}[htbp]
	\centering
	\includegraphics[width=0.7\paperwidth]{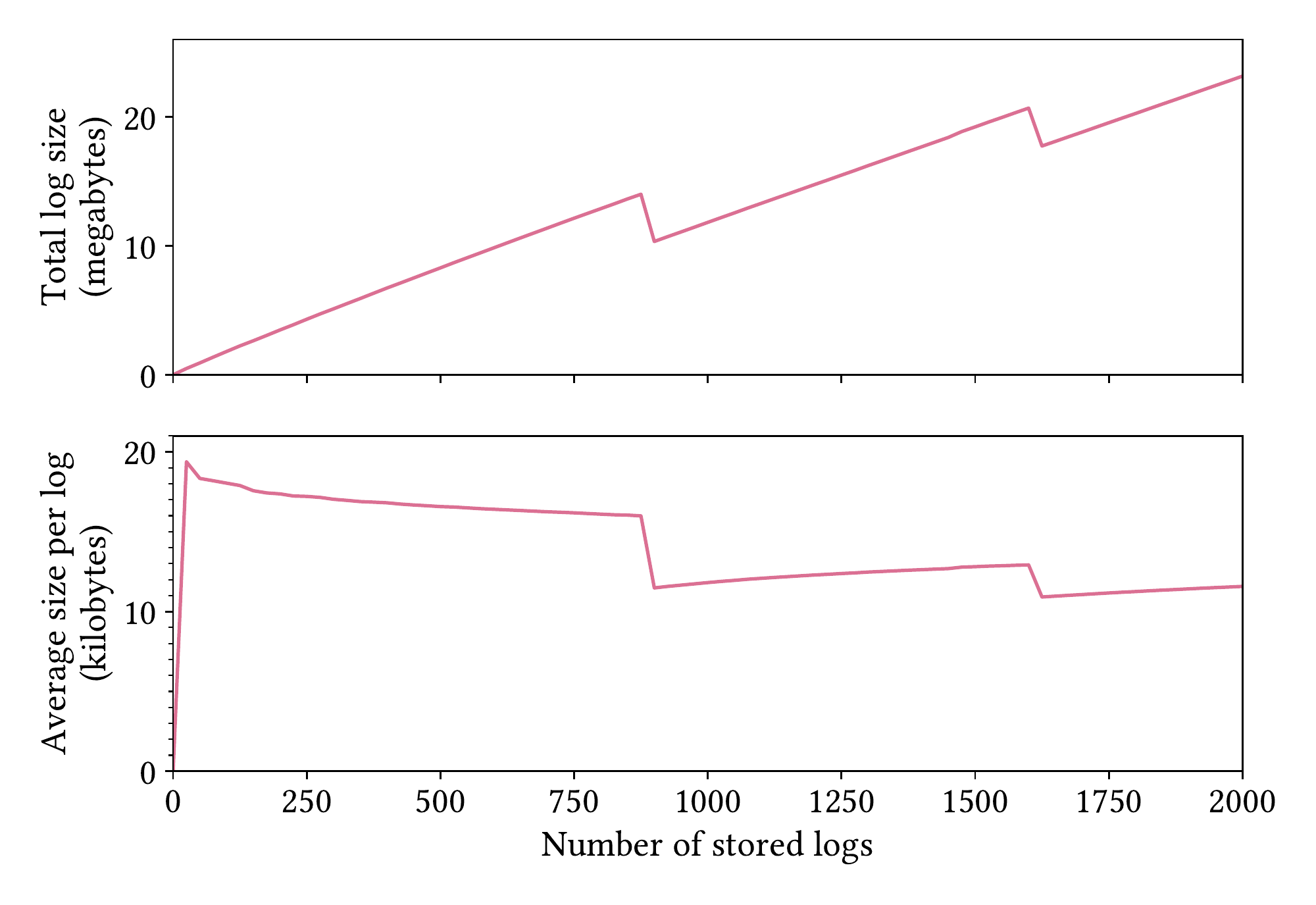}
	\caption{
		The storage requirements of the blockchain log.
		The top graph shows the total log size, which increases linearly.
		The bottom graph shows the average size per log, which decreases for larger number of logs stored in the blockchain.
		For increasing log sizes, a single entry converges to a size of 12 KB.
	}
	\Description{
		Two line graphs of the storage requirements of the blockchain.
		The top graph shows the total log size. The line increases semi-linearly, with steps, up to 24 megabytes for 2000 logs.
		The bottom graph shows the size per single entry. The line rises sharply to immediately fall again with steps, converging towards 12 kilobytes for 2000 logs.
	}
	\label{fig:evaluation:performance-storage}
\end{figure}

We find that the total log size measures in megabytes and, more importantly, rises linearly with increasing numbers of stored logs (see Fig.~\ref{fig:evaluation:performance-storage}).
For example, for 1000 logs, the total size is 11.82~MB.
With linear regression, assuming a basis storage requirement of 16 KB, we obtain a slope of approximately $12.22 \times x$~KB with $R^2 > 0.93$.
On average, an individual log entry takes up at most 20 KB, with the size approximating 12 KB per entry in a log with 2000 entries in total.
To contextualize these results, we measured the storage requirements for our usage logs when stored in a minimal SQLite database.
There, we find that storing the same log entry needs about 4.5 KB of storage space, which results in about 4.51 MB of storage for 1000 logs.
The difference is expected, as blockchain has an increased storage overhead by design.
I.e., every block includes the hash of the previous block, transaction details, and the nonce.
Importantly, though, the storage size of \Kovacs{} only linearly increases with the number of stored logs, positively impacting the scalability.

Interestingly, the blockchain size shrinks periodically (see Fig.~\ref{fig:evaluation:performance-storage}), which we attribute to Geth pruning its state~\cite{gethPruning}.
Furthermore, mining the first block increases the storage space comparatively more, which we hints at initialization overhead of the blockchain and Geth client.

\subsubsection{Conclusion}

\Kovacs{} shows good scaling behavior.
The log retrieval time rises only linearly for increasing numbers of log entries, both when retrieving a specific entry and all logs entries.
In absolute terms, it is relatively low, especially considering that no additional network requests are necessary.
Similarly, the storage requirements are, while larger than a minimal database, very low considering the additional metadata that need to be stored.
This means that storing a full blockchain copy is unproblematic for individual nodes.
We conclude that \Kovacs{} is sufficiently scalable for real-world usage and usable even in environments with hundreds of participants.

	\section{Related Work} \label{sec:related-work}

We solve two challenges of decentralized inverse transparency: non-repudiable data exchange and GDPR-compliant use of blockchain.
In the following, we discuss alternative solution approaches.

\subsection{Non-Repudiable Data Exchange}

In decentralized scenarios, achieving non-repudiation in data exchange becomes a challenge, as we have noted.
Various alternative proposals to solve this exist, of which we discuss notable examples in the following.
The overview provided by \citeauthor{wang2021staged}~\cite[Sec. 1.2]{wang2021staged} and the work by \citeauthor{kremer2002intensive}~\cite{kremer2002intensive} serve as partial foundations and help confirm our research.

\subsubsection{Protocols Requiring a Trusted Third Party}

Many data exchange protocols exist that require a trusted third party.
This includes traditional protocols based on arbitrated exchange~\cite[e.g.,][]{coffey1996non, abadi2002certified}, timing-based protocols~\cite[e.g.,][]{rabin1983transaction, zhang1996achieving}, and optimistic fair exchange protocols~\cite[e.g.,][]{asokan1997optimistic, lan2007gradual}.
This category of protocols does, by design, not fit our requirement of decentralization.
Therefore, they are on principle not relevant in our scenario.
Contrary to these approaches, we solve the issue of non-repudiable data exchange without a trusted third party.

\subsubsection{Smart Contracts as a Trusted Third Party}

To benefit from the existence of a trusted third party without necessitating the same trust, some authors propose utilizing a smart contract to fulfill that role~\cite[e.g.,][]{dziembowski2018fairswap, eckey2020optiswap}.
Compared to traditional approaches, this has the advantage that, at least theoretically, the behavior of the smart contract can be verified before it is being used.
By inspecting the smart contract code, parties can decide if they find it trustworthy.
The concrete data sharing schemes can then be modeled similarly to protocols with a regular trusted third party.
For example, they can implement arbitrated exchange~\cite[e.g.,][]{dziembowski2018fairswap} or optimistic fair exchange~\cite[e.g.,][]{eckey2020optiswap}.

Utilizing a smart contract may alleviate the issues with a trusted third party to some extent.
Compared to our solution, this approach has two main disadvantages, though.
First, it depends on the support of smart contracts by the blockchain.
Our approach, meanwhile, is compatible with any blockchain, making it more flexible.
Second, while the behavior of a smart contract can theoretically be vetted, various vulnerabilities and security issues with existing smart contracts~\cite[see, e.g.,][]{chen2020survey} show that this is a difficult problem.
In our approach, users are not expected to perform a security audit or have the technical knowledge to be able to judge the trustworthiness of a smart contract.

\subsubsection{Specialized Hardware as a Trusted Third Party}

Alternative approaches have been proposed based on trusted hardware~\cite[e.g.,][]{liu2010design, zhang2021revocable}.
Similarly to the examples with smart contracts, the specialized hardware serves as a substitute trusted third party.
For example, Intel software guard extensions~(SGX)~\cite[e.g.,][]{zhang2021revocable} or smart card~\cite[e.g.,][]{liu2010design} can be used.
Again, that means that traditional data exchange protocols can be utilized, with the trusted hardware serving as the arbiter.

Compared to using smart contracts, these approaches reduce the security and increase the required trust, though.
The utilized hardware is only considered trusted because its manufacturer is assumed to be trusted.
While with smart contracts, the actual code that is executed could---at least in theory---be vetted, trusted hardware does not even allow for this.
Additionally, this solution introduces a new problem in that only participants with the required hardware can participate.
As noted above, our approach instead does not require trust in any party.
Furthermore, it does not depend on specialized hardware.

\subsubsection{Data Delivery via Blockchain}

To not depend on any trusted entity, some authors propose to transmit the data simply by appending it to a public blockchain~\cite[e.g.,][]{zhang2021revocable}.
Immediately, the first issue with this approach becomes apparent: Potentially identifiable data are stored immutably in a blockchain.
In Sec.~\ref{sec:related-gdpr-blockchain}, we discuss why encryption is not sufficient in this case to ensure compliance with GDPR and similar privacy legislation.
Contrary to that, our approach utilizes one-time pseudonyms that guarantee unlinkability as required for anonymization.

Considering non-repudation, this approach at least ensures non-repudiation of origin, as the sending of the data is tracked in the blockchain.
Regarding non-repudiation of receipt, though, issues arise.
For example, \citeauthor{zhang2021revocable} simply claim that, due to their blockchain's inherently public nature, the receipt of data is simply ``undeniable''~\cite[p.~61]{zhang2021revocable}.
This mirrors the claim of \citeauthor{paulin2013universal}, who for their protocol claim that, as long as data are freely downloadable, their receipt can be considered as successful~\cite[p.~211]{paulin2013universal}.
We fundamentally disagree with this notion and consider it insufficient for true non-repudiation of receipt.
As the simplest example, a recipient can always claim they disconnected from the network, even after successfully receiving the data.
Accordingly, non-repudiation cannot be guaranteed in this approach, rendering it insufficient for our problem.

\subsubsection{Staged Data Delivery via Blockchain}

To generate some evidence of receipt when sharing data publicly, e.g. via blockchain, staged protocols have been proposed~\cite[e.g.,][]{paulin2013universal, wang2021staged}.
Here, the shared datum is split up into parts.
To simplify, we can generalize the solution as splitting the data up into two halves, as increasing the number of parts arbitrarily does not change the provided guarantees.
The data owner shares the first half of the encrypted data directly with the data consumer.
Then, the consumer appends an acknowledgment of receipt to the blockchain.
Only then does the owner share the second half of the data, this time via the blockchain network, with the consumer.~\cite{wang2021staged}
This improves upon full data delivery via blockchain by solving the issue of GDPR compliance.
An unreadable part of the data is also not personally identifiable and can be stored in a blockchain.

More critically, though, the receipt of the last part of the data is not acknowledged in this approach either, as in those discussed above.
Independently of how many parts of the data the recipient has acknowledged having received, if they cannot decipher the datum without receiving the last part, they can always repudiate the receipt of the full datum.
This is the fundamental issue with non-repudiation of receipt and splitting up the data does thereby not improve upon simply sending the full datum in one transaction.
Again, that means that non-repudiation cannot be guaranteed in this approach, meaning it does not represent a sufficient solution either.

\subsection{GDPR-Compliant Use of Blockchain} \label{sec:related-gdpr-blockchain}

Various proposals for how to solve the conflict between the GDPR requirements and blockchain immutability exist.
In the following, we describe important works and discuss how they differ from our approach.
The overviews by \citeauthor{pagallo2018chronicle}~\cite{pagallo2018chronicle} and \citeauthor{politou2019blockchain}~\cite{politou2019blockchain} serve as a foundation.
A recent systematic literature review~\cite{haque2021gdpr} confirms their completeness.

\subsubsection{Hashing Out}

A trivial solution would be to not store personal data in a blockchain at all.
\emph{Hashing out} specifically refers to the practice of saving only the hash of the data in the blockchain and the data themselves off-chain~\cite{pagallo2018chronicle}.
This approach is one of the most commonly used ideas to ensure GDPR compliance in blockchain solutions~\cite[see, e.g., ][]{privacyByBlockchainDesign, schaefer2019transparent, vanhoye2019logging}.
This works because the on-chain hash does not contain any private or personal data and the off-chain data can be deleted or modified to comply with a data subject's request.

There are two major downsides, though. Since the data themselves are not stored in the blockchain, this solution is not truly decentralized and requires trust in the authority managing the data~\cite{ibanez2018blockchains}.
Furthermore, using this approach one can only be sure of the \emph{existence} of entries, not of their \emph{content}.
Arbitrary entries or even the complete log could be purged, with only the hashes remaining.
To prevent malicious deletion, the party managing the log can be held accountable in case entries are missing.
This can provide some protection, but there remain options for \emph{plausible deniability}; e.g. blaming a corrupted hard disk for data loss.
That means this approach is effective only as long as the log is not tampered with, but cannot \emph{guarantee} accountability or non-repudiation.

We allow users to benefit from accountability guarantees even for highly capable adversaries---without corruptible intermediaries or plausible deniability---as we require no trusted third party.
Meanwhile, we still provide them the same level of confidentiality.

\subsubsection{Key Destruction}

If personal data \emph{are} to be stored in a blockchain, the next best idea seems to be to encrypt all stored data and delete the decryption key if the data are to be ``deleted''~\cite{pagallo2018chronicle}.

While easy to implement, this approach is flawed.
Encryption itself only guarantees pseudonymity of data~\cite{eu2016gdpr, enisa2021data}, therefore the data protection requirements still apply~\cite{enisa2021data}.
More problematically, though, if the full content of the block is encrypted, querying history becomes all but impossible, which is a requirement in secure logs for efficiently reading past entries.
The affected parties would only be able to retrieve their entries with high computational overhead, by going through every block and trying to decrypt it.

Our system in contrast enables efficient querying of entries based on the one-time pseudonyms.
The pseudonym provisioning ensures their unlinkability and enables retroactive anonymization of data, which fulfills the requirements of the GDPR's right to erasure~\cite{enisa2019recommendations, giessen2019blockchain}.

\subsubsection{Forgetting Blockchain}

\citeauthor{farshid2019design} propose to achieve a GDPR-compliant blockchain by automatically deleting blocks from the blockchain after a certain amount of time has passed~\cite{farshid2019design}.

As the described network no longer contains a genesis block, joining it becomes a challenge. The authors propose to ask other nodes for the current block and just accept it if all the returned blocks are equal~\cite{farshid2019design}. Since there is no way to verify that the received block reflects the true state of the network, joining it requires trust and does not satisfy the integrity constraint.
Secondly, the nature of their approach prevents the existence of a chain history. Applications relying on the full history, specifically in the case of secure logs, would therefore not work with this algorithm.
Furthermore, this proposal only achieves eventual GDPR compliance, since a block is only deleted after the predefined time has passed. If a user requests deletion of their data, this request cannot be fulfilled immediately.
For this reason, it is questionable if the presented idea is compatible with the GDPR.
Most problematically though, the data are only actually deleted if all nodes behave honestly~\cite{farshid2019design}. Any node can simply decide not to delete older blocks, meaning that no additional privacy guarantees can be given.

In contrast to the forgetting blockchain, our solution does not require adaptation of the utilized blockchain software and is therefore easier to integrate into existing blockchains. Furthermore, we do not depend on the honesty of \emph{arbitrary} and \emph{unknown} nodes. In contrast, only one \emph{known} party has \emph{provable} access to additional identity information and can be held liable under the GDPR.

\subsubsection{Redactable Blockchain}

The reason that the immutability of data stored in blockchain can be guaranteed is the utilized hash function:
An ideal hashing algorithm guarantees hashes that are one-way, which means impossible to reverse, and collision-free.
Then, blocks cannot be replaced without notice, as any change would result in a new hash, thereby invalidating the chain.

Redactable blockchains utilize so-called \emph{chameleon} hash functions to generate the hash of a block. Such hash functions are collision-resistant as long as a secret known as \emph{trapdoor} is not known. If one is in possession of said secret, they can efficiently compute colliding hashes~\cite{redactableBlockchain, trapdoorHash}.
With the power to create hash collisions, any block can be replaced or even removed~\cite{redactableBlockchain}, making the blockchain effectively arbitrarily editable.

In order to function, such a redactable blockchain network needs a trusted third party that is in possession of the trapdoor and can decide which block to edit~\cite{redactableBlockchain}. This constraint again requires trust, thereby calling into question the value of utilizing blockchain at all~\cite{pagallo2018chronicle}.
Furthermore, similar to the forgetting blockchain, every individual node needs to be trusted. Redactions are published as chain updates, allowing arbitrary nodes to make a copy of the removed or edited entry before updating their chain~\cite{pagallo2018chronicle, giessen2019blockchain}.
This means that, effectively, no privacy guarantees can be given.

Our solution on the contrary does not require a trusted third party and functions even in the face of adversarial network participants.

\subsubsection{Mutability by Consensus}

The introduction of a trusted third party that can arbitrarily mutate data is inherently in conflict with the core concept of blockchain.
Therefore, various proposals exist to weaken the immutability of blockchain while preserving the decentralized consensus for stored data.
Concretely, that means allowing mutations only if consensus for them is ensured.

\citeauthor{deuber2019redactable} create and formally prove an editable blockchain protocol~\cite{deuber2019redactable, politou2019blockchain}.
While any user can propose edits, the protocol ensures consensus-based voting on the proposals to prevent arbitrary edits.
This also means that no trusted third party is required.
The protocol is compatible with any consensus mechanism and even offers accountability of the performed edits.~\cite{deuber2019redactable}

While this solution removes the need for a trusted third party, it does not solve the other issue of redactable and forgetting blockchains: every individual node in the network still needs to be trusted, as mutations are published as chain updates as well.
Worse yet, the protocol introduces an additional issue in that it requires a majority of miners to act faithfully and actually perform the (legally mandated) mutations---something that it cannot guarantee by design~\cite{deuber2019redactable}.

In contrast to that, our solution functions even with of adversarial network participants, as noted above.
Furthermore, we do not depend on the honesty of the miners and, better still, do not require any changes of the blockchain software.

	\section{Limitations and Discussion} \label{sec:discussion}

Both our solution and its evaluation have limitations.
To start with, in our design, we prioritize security and decentralization.
That in turn means that other properties, such as data availability or exchange speed, are not optimized for.
Regarding data availability, each individual node manages its own data and has to be reachable when accessing data.
Should the node crash, be shut off, or otherwise disconnected from the network, the data consumer is prevented from continuing their work.
In scenarios where the availability of the nodes is prioritized higher than their security and independence, we can imagine running user's nodes, e.g., on virtual servers.
While this adds an attack surface and removes control from the user, it can improve availability.
Importantly, though, the created usage logs are highly available, as the blockchain is accessible on all nodes.
Considering exchange speed, meanwhile, we find that typical exchanges take at least 7 seconds to complete from the perspective of a data consumer.
If fake chatter is active, the median exchange duration increases by a factor of $2.4$.
In corner cases with many nodes but few exchanges, this can significantly impact scalability in the default setting.
However, in case of sufficient other traffic in the network masking the exchange, a simple heuristic could automatically deactivate fake chatter to minimize its impact.
Still, the low exchange speed is one of the largest weaknesses of our approach.
To alleviate this, requests can be pooled if more than one datum is to be requested from the same node.
Beyond that, the only other way to improve this would be to reduce the security in less critical scenarios by, e.g., introducing a name server or lowering the number of protocol rounds.
As always, this is a trade-off depending on the specific requirements.
Especially outside the workplace or if sufficient employee protection exists, less secure solutions that offer vastly higher performance may be preferable.
A sensible trade-off analysis should include security considerations but also cover factors such as cost, energy consumption, or difficulty to maintain, for example.
Yet, it is important to acknowledge that critical situations cannot always be predicted.
Therefore, we find it important to also build solutions for the most security-critical scenarios, especially if the adversity of an environment is hard to judge in advance.

Next, while our concept is fully decentralized, our implemented identity verification algorithm is built for the use case of an institutional IdP.
The IdP is, by definition, a trusted third party.
As noted in our concept (see Sec.~\ref{sec:generation-protocol}), the widely known web of trust model can be utilized instead.
We made our choice deliberately, though, as we mirror the real-world use case from industry where company-internal IdP servers are utilized for SSO.
Furthermore, implementing web of trust is in our view not a technical novelty.
Instead, we present a minimal-trust identity verification algorithm for the scenario of a company-internal IdP as a proof of concept.
To realize fully decentralized inverse transparency, an alternative identity verification scheme such as web of trust is required, though.

In our evaluation, we analyze the GDPR compliance of \Kovacs.
Due to the focus of our paper, no formal legal analysis has been performed, meaning we cannot comprehensively answer this question.
Instead, we used insights from related works to deduce the GDPR compliance of our solution.
At this moment, if and how blockchain can be used in a GDPR-compliant way has not been comprehensively answered yet, neither from a technical nor a legal perspective~\cite[see, e.g.,][]{vaneecke2018blockchain, jambert2019blockchain, tatar2020law, dePesqueraVillagran2022blockchain}.
Also, the concrete application use case is essential in conclusively determining the GDPR compliance of a solution~\cite{vaneecke2018blockchain, lyons2018blockchain}.
Therefore, before deploying \Kovacs{}, a full legal analysis including the concrete application scenario is necessary.

Furthermore, our performance and scalability evaluations are limited in their significance due to their artificial nature.
With our experiments, we tried to measure common usage scenarios and patterns.
Yet, real-world usage may differ from our tests, which can influence the performance.
As an example, a network made up of many nodes where only comparatively few nodes actually request data presents a worst case scenario for our fake chatter implementation.
The seldom communication by other nodes would require fake chatter to ensure privacy, yet the large number of potential communication partners could mean long wait times until the peer-to-peer connections are established.
The relevance of such performance bottlenecks in practice depends on the concrete usage patterns, which means real-world evaluations could be a useful next step.

Our focus was on the security of \Kovacs.
Even with the best technical protections, though, individual users remain as an often-abused attack surface~\cite[see, e.g.,][]{wl_InsiderThreat, wl_Deception, wl_Revisited, wl_Psych}.
For most data that we store, there is no danger of users unwillingly leaking information about other parties except for themselves, with one exception: \emph{data owners} could be tricked or hacked to reveal the identities of \emph{consumers} of their data.
In our current implementation, it is impossible to prevent this case, yet we consider the attack surface to be acceptably small.
To get access to a meaningful dataset about the usage pattern of a data consumer, an adversary would have to find and hack or phish each individual data owner of data accessed by said data consumer. We consider that infeasible.

Finally, to expand on these points, there has been broader discussion on what constitutes ``good enough'' software security and how to make objective judgments about it~\cite[see, e.g.,][]{sandhu2003good, dickson2011software, tondel2020achieving}.
\citeauthor{tondel2020achieving} suggest to not only consider the system from the perspective of the adversary (as we have done), but to additionally factor in other perspectives such as those of users or operators~\cite[p.~364]{tondel2020achieving}.
Following their proposal, it might therefore be sensible to conduct a broader analysis of the system that also covers these perspectives before deploying it.
This could be important to ensure user acceptance and usability, as well as to address potential practical issues with deployment and operation that might otherwise hinder adoption.

	\section{Conclusion}

The goal of inverse transparency is to protect employees from misusage of their data.
Yet, current technical realizations are inherently centralized, which requires trust and opens possibilities for tampering with the logs by, e.g., the employer.
Permissionless blockchain therefore is an intuitive choice for inverse transparency logs, as it is by design decentralized and immutable.
Realizing fully decentralized inverse transparency with blockchain requires us to tackle two main issues, though:
(1)~ensuring non-repudiable data exchanges without a trusted third party and
(2)~complying with GDPR requirements, specifically confidentiality and the right to erasure.
With the \Kovacs{} system, we solve both of these issues.
For accountable inverse transparency, its new-usage protocol enables decentralized and non-repudiable data exchange.
To enable GDPR compliance, its pseudonym generation algorithm guarantees unlinkability and anonymity of stored data, while enabling proof of ownership and authenticity.
Our block structure and decentralized deployment architecture allow individuals to efficiently query for and read arbitrary usage log entries concerning their data, while protecting them from attacks on their confidentiality by adversaries.

In our analysis, we find that \Kovacs{} provides a high level of security and protects against expected attacks on the confidentiality of the logs.
It fulfills the requirements of the GDPR by enabling confidentiality and the rights to erasure and rectification, while at the same time benefiting from the properties of permissionless blockchain, specifically guaranteeing the integrity of the logged data.
Related works require either the use of a permissioned blockchain, necessitating a trusted third party, or modifying the utilized hashing algorithms or blockchain software to make the blockchain mutable. Both approaches entail effectively giving up the advantages of blockchain, thereby calling into question the use of blockchain in the first place.
Our performance and scalability evaluations demonstrate the practicality of our implementation.
While our focus on security impacts performance, the exchange duration and query times stay manageable for typical workloads.
If exchange speed is prioritized, our protocol can be adapted flexibly.
Furthermore, we find that \Kovacs{} scales linearly considering both the retrieval time and storage size, showing its practicality.
The additional metadata stored mean higher storage requirements than a minimal database, but the logs are still sufficiently small.

To conclude, the \Kovacs{} system realizes decentralized, non-repudiable, secure, and GDPR-compliant inverse transparency based on blockchain.
Its design does not require a trusted third party, it can be used with any existing blockchain software without necessitating changes, and it is secure and flexible enough for integration even into highly adversarial settings.


	\begin{acks}
		This work was supported by the \grantsponsor{BMBF}{German Federal Ministry of Education and Research (BMBF)}{https://www.bmbf.de/} under grant no. \grantnum{BMBF}{5091121}.
	\end{acks}

	\bibliographystyle{ACM-Reference-Format}
	\bibliography{references}


\end{document}